\documentclass[aps,prl,a4paper,notitlepage,twocolumn,noshowpacs,longbibliography,10pt]{revtex4-1}
\usepackage[utf8]{inputenc}
\usepackage[T1]{fontenc}
\usepackage{lmodern}
\usepackage{amsfonts}
\usepackage{amssymb}
\usepackage{bm} 
\usepackage{graphicx}
\usepackage{isomath}
\usepackage{xcolor}
\usepackage{siunitx}
\usepackage{amsmath}
\usepackage[version=3]{mhchem}

\begin{document}
\title{Vortex-mediated relaxation of magnon BEC into light Higgs quasiparticles}

\author{S.~Autti$^{1,2\ast}$}

\author{P.J. Heikkinen$^{1,3}$}

\author{S.M. Laine$^4$}

\author{J.T. M\"akinen$^{1,5,6}$}
\author{E.V. Thuneberg$^{4,7}$}

\author{V.V. Zavjalov$^{1,2}$}

\author{V.B.~Eltsov$^{1}$}
\affiliation{$^1$Low Temperature Laboratory, Department of Applied Physics, Aalto University, POB 15100, FI-00076 AALTO, Finland. \\
$^2$Department of Physics, Lancaster University, Lancaster, LA1 4YB, UK. *Email: s.autti@lancaster.ac.uk\\
$^3$Department of Physics, Royal Holloway, University of London, Egham, Surrey, TW20 0EX, UK. \\
$^4$Nano and Molecular Systems Research Unit, University of Oulu, P.O. Box 3000, Oulu FI-90014, Finland \\
$^5$Department of Physics, Yale University, New Haven, CT 06520, USA\\
$^6$Yale Quantum Institute, Yale University, New Haven, CT 06520, USA\\
$^7$QTF Centre of Excellence, Department of Applied Physics, Aalto University, FI-00076 AALTO, Finland. 
}

\begin{abstract}
A magnon Bose-Einstein condensate in superfluid $^3$He is a fine instrument for studying the surrounding macroscopic quantum system. At zero temperature, the BEC is subject to a few, distinct forms of decay into other collective excitations, owing to momentum and energy conservation in a quantum vacuum. We study the vortex-Higgs mechanism: the vortices relax the requirement for momentum conservation, allowing the optical magnons of the BEC to transform into light Higgs quasiparticles. This observation expands the spectrum of possible interactions between magnetic quasiparticles in $^3$He-B, opens pathways for hunting down elusive phenomena such as the Kelvin wave cascade or bound Majorana fermions, and lays groundwork for building magnon-based quantum devices.
\end{abstract}

\maketitle

One illuminating perspective to the ground state of a fermionic condensate, such as zero-temperature superfluid $^3$He, is to treat it as a quantum vacuum where moving objects interact with the excitations of the vacuum \cite{VolovikBook,bradley2016breaking,PhysRevB.98.144512,PhysRevResearch.2.033013,autti2020fundamental}. Various collective excitations, for example magnetic quasiparticles (magnons), and topological defects such as quantised vortices can be manipulated in this extremely pure environment. A Bose-Einstein condensate of optical magnons (magnon BEC), trapped within the superfluid, can be instrumented to probe objects in the system without influencing them \cite{1992_ppd,2000_ppd,magnon_trap_mod,PhysRevLett.121.025303}. This capacity has inspired suggestions to use the BEC to detect surface- or vortex-core-bound Majorana fermions \cite{Muarakawa2011,Shapiro2013} or the Kelvin wave cascade \cite{eltsov2020amplitude,l2010spectrum}. Both have so far remained elusive despite decades of active 
research. Changes in the BEC ground state frequency as well as the population decay rate of the BEC can be devised for such purposes, provided the basic interactions between the excitations of the quantum vacuum are first thoroughly mastered. On the other hand, macroscopic quantum systems such as BEC-based time crystals\cite{PhysRevLett.120.215301,PhysRevB.100.020406,autti2020ac} provide a promising building block for quantum technologies, which rely on controlled non-destructive manipulation of the system. Such control can be accessed in the superfluid vacuum by coupling the BEC to and decoupling it from available excitations selectively.

In superfluid $^3$He, the spin and orbital angular momenta of Cooper pairs are equal to one. In the B phase, the relative spin-orbit symmetry is broken in addition to the emergence of a coherent phase, as described by a $3 \times 3$ complex order-parameter matrix \cite{leggett1975theoretical,vollhardt1990}. The macroscopic spin and orbital momentum directions are connected by the spin-orbit rotation angle $\theta_\mathrm{L}\approx104^\circ$ around axis $\hat{{\bm n}}$. The fermionic thermal excitations of this system have energy gap $\Delta_\mathrm{B}$ which is on the order of $k_\mathrm{B} T_\mathrm{c}$, where $k_\mathrm{B}$ is the Boltzmann constant and $T_\mathrm{c}$ the superfluid transition temperature. At temperatures much below $T_\mathrm{c}$, the number of thermal excitations is reduced exponentially, creating a vacuum void of fermionic quasiparticles.

\begin{figure}[tb]
\centering
\includegraphics[width=1\columnwidth]{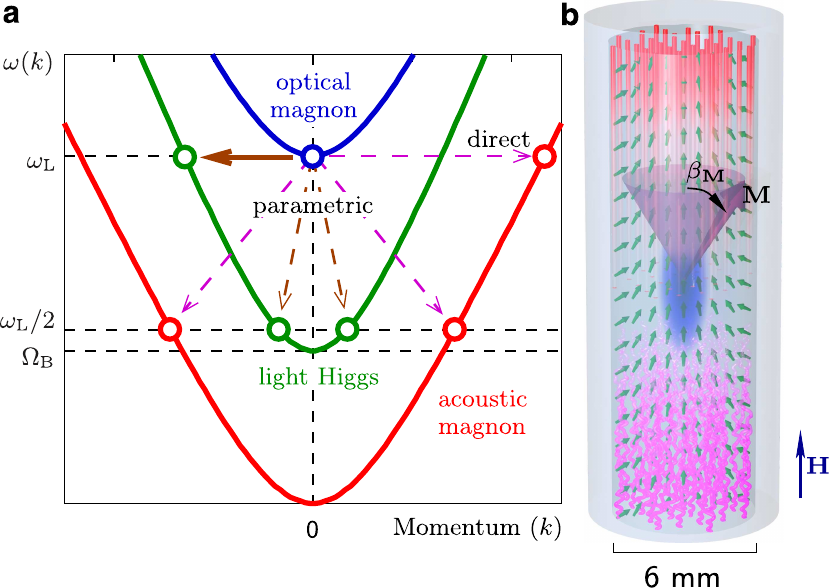}
\caption{({\bf a}) Spectra of spin waves in $^3$He-B. The mass (gap) $\omega_\mathrm{L}$ of optical magnons (blue line) can be tuned using the magnetic field. Direct conversion of optical magnons into light Higgs quasiparticles (solid arrow), studied in this Letter, requires balancing the momentum mismatch. The parametric conversion of optical magnons into gapless acoustic magnons (red line), and light Higgs quasiparticles (green line) can be observed when the density and mass of the optical magnons is large. Processes indicated by dashed arrows were reported in Ref.~\onlinecite{HiggsNComm}. ({\bf b}) Superfluid $^3$He in a cylindrical container. A BEC of optical magnons (blue blob) is trapped in the middle by the spatial distribution of the orbital order parameter ($\bm{\hat{n}}$-vector, small green arrows) and by an axial minimum in the external magnetic field ${\bm H}$. The coherently precessing magnetisation ${\bm M}$ (large magenta arrow) in the BEC is parametrised with the tipping angle $\beta_\mathrm{
M}$. In constant rotation $\Omega$ around the vertical axis, an array of vortices is created, penetrating the BEC (red vertical rods at the top). For illustrational reasons the vortex rods have been made transparent in the vicinity of the BEC, and drawn only in the upper half of the container. The vortex configuration obtained in modulated rotation is sketched with the magenta rods at the bottom of the container, based on Ref.\ \cite{PhysRevB.97.014527}.}
\label{3He_modes}
\end{figure}

Besides the fermionic quasiparticles, there are three collective spin-wave modes with a small (or zero) gap, corresponding to the combined oscillations of three spin components and three components of spin-orbit rotation \cite{vollhardt1990}. Following Ref.~\onlinecite{HiggsNComm}, we call these modes optical magnons, acoustic magnons, and light Higgs quasiparticles. In the absence of a magnetic field, optical and acoustic magnons are gapless, corresponding to the oscillations of $\hat{{\bm n}}$. That is, their frequency vanishes in the long wave length limit. The light Higgs mode corresponds to oscillation of the spin-orbit rotation angle around its equilibrium value $\theta_\mathrm{L}$, 
 and has a gap $\Omega_\mathrm{B}/2\pi\sim 100$~kHz ($\Omega_\mathrm{B}$ is the Leggett frequency).  In a magnetic field $H$, optical magnons acquire a gap equal to the Larmor frequency $2\pi f_\mathrm{L}=\omega_\mathrm{L}=\gamma H$, where $\gamma$ is the gyromagnetic ratio. The dispersion relations of the three modes are illustrated in Fig.~\ref{3He_modes}(a). 

The superfluid vacuum can also host topological defects \cite{VolovikBook}, in particular quantised vortices. An ordered array of vortices can be created by rotating the sample at a constant angular velocity $\Omega$. The density of the vortex array is proportional to $\Omega$. The B-phase vortices have a broken-symmetry core \cite{RevModPhys.59.533,lounasmaa1999vortices}, where the low-temperature vortex studied in this work has a double-core structure consisting of two tightly-bound sub cores \cite{thuneberg1987,volovik1990half,fogelstrom1995,silaev2015,kasamatsu2019effects,nagamura2019doublecore,regan2020vortex}. A theoretical description of the interactions of vortices and collective excitations in superfluid $^3$He can be confidently constructed by expanding on the BCS theory with Fermi-liquid effects included \cite{serene1983quasiclassical}.

In this Letter we study the interaction of a condensate of optical magnons with quantised vortices and light Higgs quasiparticles, which we call the vortex-Higgs mechanism: If a vortex penetrates the magnon BEC, optical magnons are scattered by the order-parameter distortion that surrounds the vortex. This interaction lifts the requirement for momentum conservation for inbound and outbound quasiparticles. We show that in such collisions the optical magnons in the condensate are converted directly into light Higgs quasiparticles. This is seen as zero-temperature relaxation of the BEC with exponential time dependence. We study this conversion in two qualitatively different vortex configurations, ordered and disordered, and find that the results are in good agreement with theory.

The magnon BEC in superfluid $^3$He consists of coherent optical magnons \cite{magnon_BEC_review}. Their magnetisation ${\bm M}$ precesses around the external magnetic field ${\bm H}$ and is described by a macroscopic wave function $\Psi$. The total number of magnons  $N\propto\int|\Psi|^2\mathrm{d}V\propto \int\beta_{\bm M}^2\mathrm{d}V$, where $\beta_{\bm M}$ is the deflection angle of ${\bm M}$ from the equilibrium direction along ${\bm H}$, and $V$ is volume. Here we assumed that $\beta_{\bm M}$ is small, which is satisfied in all the experiments presented in this Letter. The coherently precessing magnetization  is generated and detected using Nuclear Magnetic Resonance techniques.

The magnon BEC is trapped in the middle of the superfluid sample (Fig.~\ref{3He_modes}b) by the combined effect of the orbital order parameter distribution (``texture''), and a profile of the external magnetic field. The resulting trap is nearly harmonic\cite{magnon_relax,zavjalov2015measurements}, characterised by the radial and axial trapping frequencies, $f_\mathrm{r}$ and $f_\mathrm{z}$, determined from measurements of the full spectrum of states in the trap \cite{zavjalov2015measurements}. We concentrate on the ground state magnon BEC, whose precession frequency is $f=f_\mathrm{L}+f_\mathrm{r} + f_\mathrm{z}/2$.  Temperature is measured using a mechanical oscillator, a quartz tuning fork. Its resonance width follows $\Delta \nu\propto \exp(-\Delta_\mathrm{B}/k_\mathrm{B} T)$ at temperatures $T\ll T_\mathrm{c}\sim 1~$mK, probing the density of thermal quasiparticles \cite{2007_forks}. Details of the experimental setup can be found in Refs.~\cite{magnon_relax,Heikkinen2014}.

\begin{figure}[tb!]
\centering
\includegraphics[width=1\linewidth]{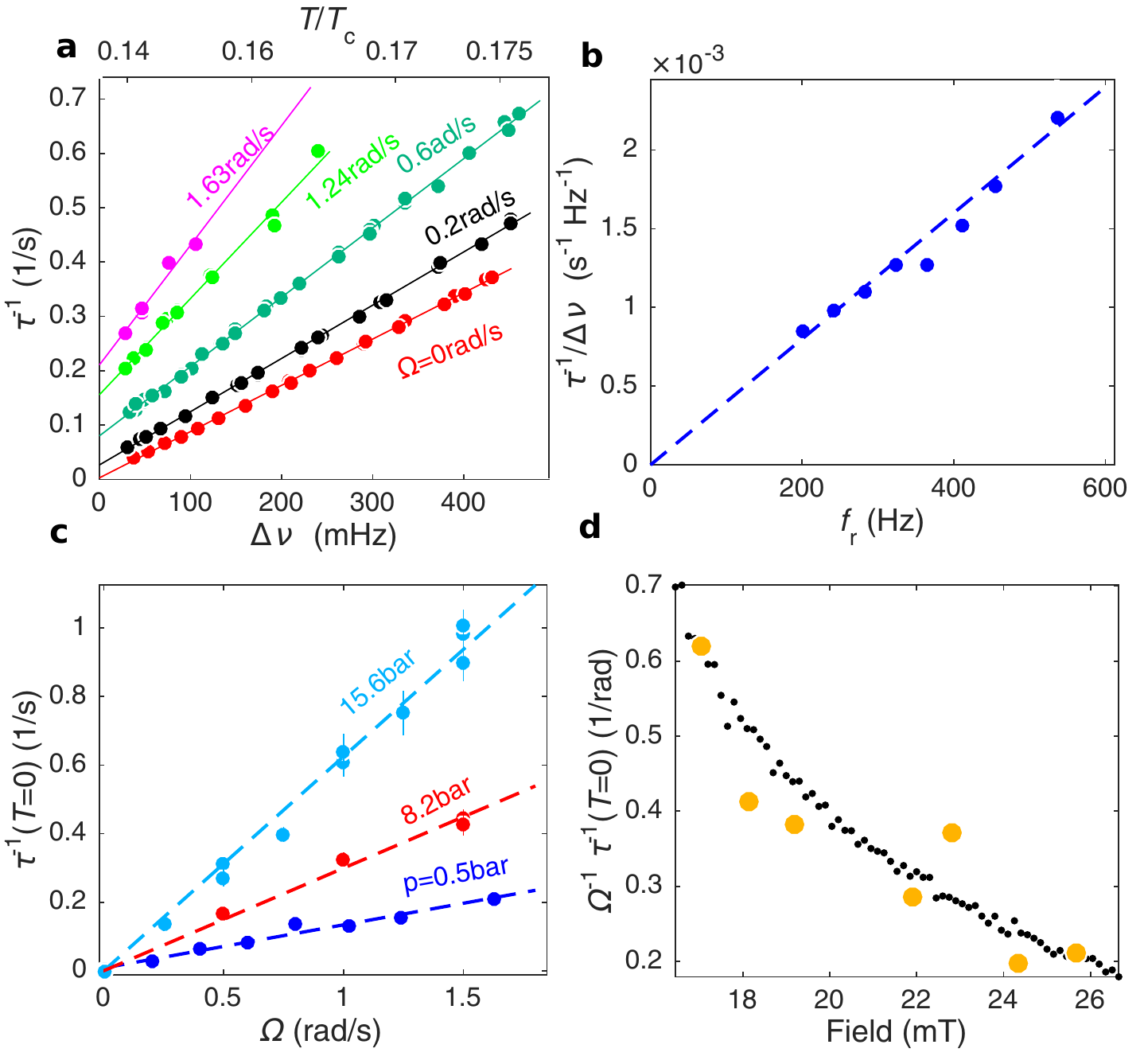}
\caption{Vortex-Higgs mechanism in steady rotation: ({\bf a}) Measured relaxation rate at $\Omega=0$ (red points) is linear in the thermometer fork width $\Delta \nu$ owing to spin diffusion (red line). Intrinsic fork width has been subtracted from $\Delta \nu$ shown here. Data measured at $\Omega>0$ shows an increased slope, reflecting the changing trap shape. ({\bf b}) The temperature-dependence slopes, $\frac{\mathrm{d}\tau^{-1}(\Omega)}{\mathrm{d}\Delta \nu}$,  in panel {\bf a} (points) are proportional to the radial trapping frequency $f_\mathrm{r}$ (dash line is a linear fit through zero), as expected for spin diffusion, implying that other relaxation contributions are temperature-independent. ({\bf c})  The temperature-independent relaxation extracted as illustrated in panel {\bf a} by extrapolating to $\Delta \nu=0$ (coloured dots) is proportional to $\Omega$, that is, to the vortex density (dash lines).  ({\bf d}) The magnetic field dependence of the vortex relaxation in the ordered state, extracted 
similarly as shown in panel {\bf a} (large orange circles, $\Omega=1~\mathrm{rad\, s^{-1}}$), is in good agreement 
with that obtained by modulated rotation (small black points).  Dissipation due to losses in the measurement circuity and the intrinsic fork width have been subtracted from all data as explained  in Ref.~\onlinecite{magnon_relax} and Fig.~\ref{Modulation1}. Pressure in panels {\bf a} and {\bf b} was 0.5~bar. The magnetic field for the $0.5~$bar and 15~bar data corresponds to $f_\mathrm{L}=826~$kHz, and for the 4~bar and 8~bar data to $f_\mathrm{L}=833~$kHz. Data in panel {\bf d} was measured at 4~bar pressure. Error bars correspond to uncertainty in removing the resonant relaxation peaks.}
\label{relax1}
\end{figure}

At a finite temperature, the relaxation of a magnon BEC is primarily caused by non-hydrodynamical spin diffusion \cite{magnon_relax,Heikkinen2014}, resulting in exponential decay of the condensate, $\beta_{\bm M} \propto \exp(-t/\tau_\mathrm{SD})$. Here $1/\tau_\mathrm{SD}$ is the spin-diffusion relaxation rate. Spin diffusion is proportional to the thermal quasiparticle density, leading to a linear dependence between the thermometer fork resonance width and the BEC relaxation rate, $1/\tau_\mathrm{SD} \propto \Delta \nu$ (Fig.~\ref{relax1}a) \cite{magnon_relax}.  In practice there are also unavoidable losses in the measurement circuitry, but this effect can be confidently subtracted \cite{magnon_relax}. 

In the zero-temperature limit  intrinsic decay channels are absent and the condensate lifetime approaches infinity \cite{2000_ppd,PhysRevB.86.024506,PhysRevLett.120.215301,autti2020ac}. Any (extrapolated) zero-temperature dissipation in the bulk liquid \cite{PhysRevB.86.024506} is therefore an indication of interaction with other collective modes either via parametric excitation or direct conversion. The former is allowed assuming the density of optical magnons is high enough and their mass large enough \cite{HiggsNComm}. Direct conversion is ruled out due to momentum conservation unless mediated by boundaries or interaction with topological defects of the superfluid vacuum, such as quantum vortices. The interaction of magnons with a topological defect arises due to the distortion of the order parameter distribution in the vicinity of the defect. For a vortex this is quantified by a coupling constant $C$, which gives the amplitude of the deviation of the spin-orbit rotation from the equilibrium. This 
deviation decays as one over the distance from the vortex line. For the double-core vortex, $C$ is nearly equal to the separation of the half cores. The coupling mechanism is derived in Ref.~\onlinecite{laine2018} and applied for the present case in Supplementary Note.

The entire refrigerator used in the experiments can be rotated around the axis of the sample container cylinder. Rotation creates an equilibrium array of vortices, which has a twofold effect on the BEC. First, the global orbital texture reacts to the vortex array \cite{eltsov2011vortex}, changing the shape of the trap ($f_\mathrm{r}$ increases). This affects the spin diffusion relaxation, which can be written as $1/\tau_\mathrm {SD}= f_\mathrm{r} D \times \mathrm{const.}$ \cite{Heikkinen2014} ($f_\mathrm{r} \gg f_\mathrm{z}$). Here $D$ is the applicable component of the transverse spin diffusion tensor. We find that changes in the measured relaxation $1/\tau$ are proportional to $\Delta \nu$ at any given $\Omega$ (Fig.~\ref{relax1}a), and that the slope $\frac{\mathrm{d}\tau^{-1}(\Omega)}{\mathrm{d}\Delta \nu}$ is proportional to the measured radial trapping frequency $f_\mathrm{r}$ (Fig.~\ref{relax1}b). This observation implies that the temperature dependence of the relaxation rate $1/\tau$, contained in 
$D$, is not affected by rotation, and any relaxation directly related to the vortices is temperature-independent below $T=0.17 T_\mathrm{c}$. We emphasise that all the relaxation signals measured were exponential in time, implying that no non-exponential contribution was added by the vortex array. The second observation is that the zero-temperature ($\Delta \nu=0$) extrapolation of the relaxation is proportional to $\Omega$ (Fig.~\ref{relax1}c). That is, the observed temperature-independent relaxation is (i) also exponential and (ii) proportional to the density of vortices. This is in good agreement with the theoretical expectation for vortex-Higgs mechanism of BEC relaxation, Eq. (\ref{final_relax}), 

We note that peaks in the measured relaxation, associated with the presence of vortices, were observed with roughly 1~kHz spacing in $f_\mathrm{L}$ on top of the vortex-Higgs dissipation described above. We account this phenomenon for resonant production of standing spin wave modes in the sample container, mediated by the vortex array \cite{HiggsNComm}. Both acoustic magnons and light Higgs quasiparticles are viable candidates to explain this observation, but a detailed study is left for a future publication. For simplicity, in what follows we call the peaks ``relaxation peaks''. The peak frequencies were avoided in all measurements conducted at stable rotation.

As an alternative to constant rotation, the angular velocity can be modulated.
We used linear modulation in the range from 1.4~$\mathrm{rad\, s^{-1}}$to 1.8~$\mathrm{rad\, s^{-1}}$ with $|\mathrm{d}{\Omega}/\mathrm{d}t|=0.03~\mathrm{rad\, s^{-2}}$. In the steady state the vortex number is expected to remain constant but the vortex array is distorted. We find experimentally that this removes the resonance peaks found at constant rotation. 
The measured zero-temperature relaxation with and without modulation, the latter avoiding the relaxation peaks, are shown in Fig.~\ref{relax1}d. The two vortex configurations yield the same BEC relaxation rate. This observation allows us to probe the vortex-Higgs mechanism at arbitrary magnetic fields, avoiding the relaxation peaks altogether. 

\begin{figure}[tb!]
\centering
\includegraphics[width=1\columnwidth]{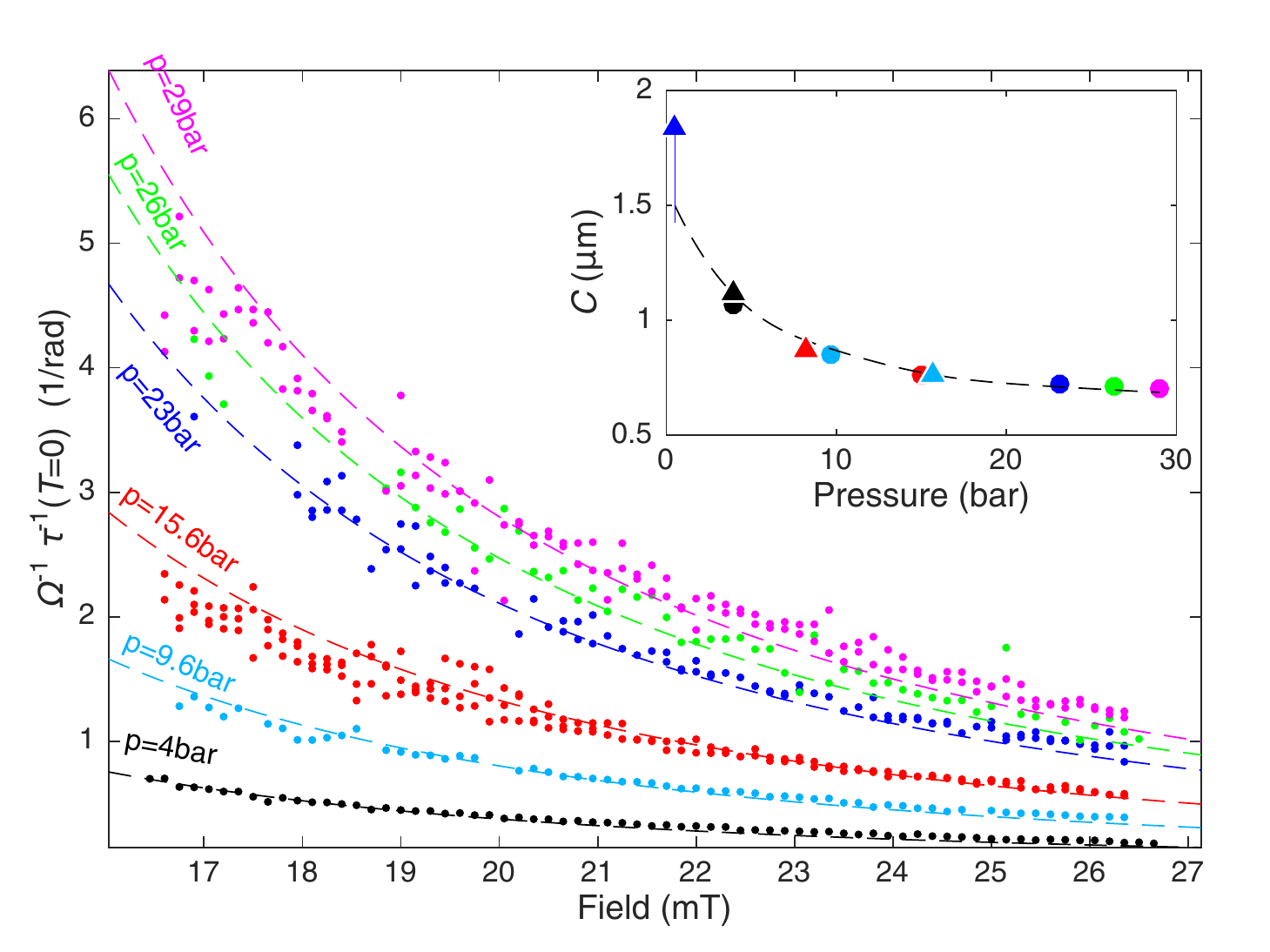}
\caption{Vortex-Higgs mechanism as a function of magnetic field and pressure: Measured BEC-relaxation field dependence at different pressures (coloured dots) is in good agreement with the theoretical expectation (dash lines) for vortex-mediated conversion of BEC magnons into light Higgs quasiparticles. Theory lines correspond to Eq.~\ref{final_relax}, fitted to the data using parameter $C$. All measurements were carried out at $T=0.15 T_\mathrm{c}$. Spin diffusion dissipation and radiation losses have been subtracted based on measured trapping frequencies $f_\mathrm{r}$ and $f_\mathrm{z}$ \cite{Heikkinen2014}. This correction is about 5 \% for the 4~bar data, and less than 1\% for the 29~bar data. The inset shows $C$ vs. pressure, in good agreement with theoretical expectation $C=\tilde{C} R_0$, with fitted $\tilde{C}=7.4$ (dashed line). The coloured circles correspond to fits in the main panel, and the triangles to the straight-vortex data in in Fig.~\ref{relax1}~c,d. Error bars correspond to uncertainty in 
removing the resonant relaxation peaks.}
\label{Modulation1}
\end{figure}

We can further characterise the vortex-Higgs mechanism by varying the coupling between magnons and the vortices. The coupling constant $C$ can be controlled with the pressure and the magnetic field, as derived in the Supplementary Note. We compare the experiment with the theoretical model in  Fig.~\ref{Modulation1}. The data is fitted using the coupling constant $C$ (other parameters were taken from Ref.~\onlinecite{thuneberg2001}). We find very good agreement in the magnetic field as well as the pressure dependence with $C=7.4 R_0$ ($R_0$ is a characteristic length scale of the order of the coherence length in the superfluid as defined in the Supplementary Note). This value is close to the theoretical value $C/R_0=5.9-6.6$ \cite{silaev2015}, and therefore the agreement between experiment and theory is highly satisfactory without any fitting parameters. That is, the millimetre-sized magnon BEC correctly measured the effective vortex half-cores' separation which is of the order one micrometre, confirming the 
assumption that the BEC interacts with each individual vortex independently and providing a strong argument in support of the vortex-Higgs interpretation. 

The observations presented above imply that the vortex-Higgs mechanism opens an otherwise-unavailable relaxation channel for the magnon BEC, corresponding to zero-temperature conversion of optical magnons of the BEC into light Higgs quasiparticles. This connection is mediated by the double-core vortex of low-temperature superfluid $^3$He, and it is robust against changes in vortex orientation and order. The role of the vortices, acting via the textural distortion that surrounds them, is to bridge the mismatch of momentum between the two species of quasiparticles. The measured dissipation is proportional to vortex density, and reacts to changes in the core dimensions of the B phase vortices and the order parameter distribution surrounding them as expected: the measured magnetic field and pressure dependencies as well as the temperature independence of the measured relaxation follow the theory developed in the Supplementary Note. These observations add the vortex-Higgs mechanism to the set of confirmed 
interaction channels of magnetic quasiparticles in superfluid $^3$He. 

It remains an interesting task for the future to confirm further predictions of the vortex-Higgs mechanism. The conversion of optical magnons into light Higgs quasiparticles is only possible via this mechanism assuming the optical magnons have a mass ($\propto\omega_\mathrm{L}$) larger than that of light Higgs quasiparticles ($\propto\Omega_\mathrm{B}$). The former is controlled by the magnetic field, and the latter by pressure. Probing the region where $\omega_\mathrm{L}<\Omega_\mathrm{B}$, and thus the vortex-Higgs mechanism is disabled, requires developing a spectrometer capable of measuring at a sufficiently low magnetic field (<8~mT). Such setup could be used to conclusively identify the source of the resonant relaxation peaks as well. On the other hand, our results provide basis for looking for relaxation contributions beyond the vortex-Higgs mechanism, for example those originating from vortex-core- or surface-bound fermionic quasiparticles \cite{Muarakawa2011,Shapiro2013,PhysRevLett.108.045303,
PhysRevB.44.9667,kopnin1998rotating}. For such studies one should also operate below the cut-off frequency $\Omega_\mathrm{B}$, allowing the detection of smaller relaxation contributions.

The magnon BEC also makes a sophisticated new tool for probing other emergent phenomena, such as vortex dynamics. In particular, zero-temperature vortex turbulence is believed to be terminated by the Kelvin wave cascade, but it remains a long-standing challenge to confirm and explore this effect experimentally \cite{eltsov2020amplitude,l2010spectrum,kozik2004kelvin,PhysRevLett.86.3080}. On the other hand, surfaces and vortex cores of superfluid $^3$He host elusive Majorana bound states \cite{Muarakawa2011,Shapiro2013}, with a characteristic zero-temperature dissipation signature \cite{chung_prl103}, which remains to be conclusively evidenced. Both these realms can now be explored using a magnon BEC as an instrument. For example, one way of detecting the Majorana quasiparticles relies on changing the relative magnetic field orientation and measuring the related magnetic relaxation \cite{chung_prl103}. Our work provides a solid basis for measuring any delicate field dependence of the BEC relaxation rate. 
Finally, applying the BEC to build quantum devices \cite{autti2020ac} --- eventually perhaps even at room temperature \cite{alex2019josephson,Bozhko2019,PhysRevB.100.020406,PhysRevLett.121.077203,Bozhko20161057} --- is an exciting new avenue for research that our work and other recent advances have enabled. Quantum vortices could be used to manipulate the state and features of such a device via the vortex-Higgs mechanism with minimal disturbance and coupling to the external world.

This work has been supported by the European Union's Horizon 2020 research and innovation programme (grant no. 694248). The experimental work was carried out in the Low Temperature Laboratory, which is a part of the OtaNano research infrastructure of Aalto University and of the EU H2020 European Microkelvin Platform (grant No. 824109). S.A. and V.V.Z. were funded by UK EPSRC (grant no.EP/P024203/1). S.A. acknowledges support from the Jenny and Antti Wihuri foundation, P.J.H. that from the V\"{a}is\"{a}l\"{a} foundation of the Finnish Academy of Science and Letters, and S.M.L. that from both of the above. E.V.T. acknowledges support by the Academy of Finland Centre of Excellence program (project 312057). 

%

\clearpage
\newpage
\onecolumngrid

\renewcommand{\thefigure}{S\arabic{figure}} 
\setcounter{figure}{0}
\renewcommand{\theequation}{S\arabic{equation}} 
\setcounter{equation}{0}

\clearpage
\newpage

\section*{Supplementary Note}

In order to derive a description of the vortex-Higgs mechanism,
let us study the interaction between one double-core vortex and coherently precessing magnetisation. To make the calculations tractable, we ignore the presence of the container and other vortices. We follow closely the theory presented in Ref.~\onlinecite{laine2018}, although with slightly different notation.

The geometry of the problem is as follows. The vortex line is aligned along the $z$ axis, and is oriented in the $x$-$y$ plane so that the two sub cores lie along the $y$ axis. We assume that there is an external static magnetic field pointing along $\hat{{\bm z}}$. We also assume that the unperturbed magnetisation precesses uniformly about $\hat{{\bm z}}$ with a small tipping angle $\beta_{\bm  M}$, as in the experiment presented in the main text of this Article, and work in lowest non-trivial order in $\beta_{\bm  M}$.

Outside the immediate core region of the vortex, the order parameter of the system is proportional to a rotation matrix,
\begin{equation}
\mathsf{A} \propto \mathsf{R} \left( \bm \theta_0 \right) \mathsf{R} \left( \bm \theta_1 + \bm \theta_2  \right).
\end{equation}
Three factors affect the direction and the angle of the rotation. One is the magnon condensate, which fixes the vector $\bm \theta_0=\theta_L\hat{{\bm n}}$, where the rotation angle $\theta_L =  \arccos(-1 / 4)$ and the rotation axis $\hat{{\bm n}} = \hat{{\bm z}} + \sqrt{2/5} \beta_{\bm  M} \mathsf{R}(\hat{{\bm z}}\omega_\mathrm{L} t) \cdot \hat{{\bm y}}$.  The vortex modifies this by the asymptotic form
\begin{equation}\label{e.asymp}
\bm \theta_1 = \frac{C \cos \varphi}{r} \left( \frac{\sin \varphi}{1+c} \hat{{\bm r}} + \cos \varphi \bm{\hat{\varphi}} \right).
\end{equation}
Here $(r, \varphi)$ are the standard polar coordinates so that $x=r \cos \varphi  $ and $y=r \sin \varphi $, and parameter $c \sim 1$ depends on temperature and pressure. The coefficient $C$ is approximately equal to the distance between the sub cores of the vortex, and also depends on temperature and pressure. Finally, there is a contribution $\bm \theta_2$ stemming from the interaction between the condensate and the vortex. It satisfies a wave-like equation
\begin{equation}\label{eq:eom}
\begin{split}
&\bm{\ddot{\theta}}_2 - \omega_\mathrm{L} \hat{{\bm z}} \times \bm{\dot{\theta}}_2 + \Omega_B^2 \bm{\hat{z}} \left(\hat{{\bm z}} \cdot \bm \theta_2 \right) - v^2 \left[ (1+c) \nabla^2 \bm{\theta}_2 - c \nabla \left( \nabla \cdot \bm{\theta}_2 \right) \right] = \bm \rho
\end{split}
\end{equation}
together with the boundary condition $\lim_{r \to 0} \bm \theta_2 = 0$ at the origin and a causal boundary condition at $r \to \infty$. Here, $v = 2 \xi_D \Omega_B / \sqrt{15}$, $\xi_D$ is the dipole length,
\begin{align}
\bm \rho (\bm r, t) &= \Re \left\{ e^{-i \omega_\mathrm{L} t} \tilde{\rho}(r, \varphi) \right\} \hat{{\bm z}}, \label{eq:rho_vector} \\
\tilde{\rho}(r, \varphi) &= -\sqrt{\frac{2}{5}} \frac{\beta C \Omega_B^2}{r} \cos \varphi \frac{2 + c + c e^{2 i \varphi}}{2(1 + c)},
\end{align}
and $\Re\{\cdot\}$ denotes the real part of a complex number.

The deviation of the magnetisation from the coherently precessing one is related to the time derivative of $\bm \theta_2$ by
\begin{equation}
\delta {\bm  M} = \frac{\chi}{\mu_0 \gamma} \mathsf{R}_z(\theta_L) \cdot \bm{\dot{\theta}}_2,
\end{equation}
where $\chi$ is the B-phase susceptibility, $\mu_0$ is the vacuum permeability, and $\gamma$ is the gyromagnetic ratio of $^3$He.

\subsection*{Spin wave modes}
In order to gain more insight into the equation of motion (\ref{eq:eom}), let us consider the eigenmodes $\bm \theta_2(\bm r, t) = {\bm A} e^{i({\bm k} \cdot {\bm r} - \omega t)}$ of the corresponding homogeneous equation. Since the source term $\bm \rho$ in Eq.\ (\ref{eq:eom}) is independent of $z$, we are only interested in modes with ${\bm  k} \cdot \hat{{\bm z}} = 0$. Substituting the ansatz into the equation yields
\begin{equation}
-\omega^2 {\bm A} + i \omega_\mathrm{L} \omega \hat{{\bm z}} \times {\bm A} + \Omega_B^2 \hat{{\bm z}}(\hat{{\bm z}} \cdot {\bm A}) + v^2 \left[ (1 + c) k^2 {\bm A} - c {\bm  k}({\bm  k} \cdot {\bm A}) \right] = \bm 0.
\end{equation}
There are two transverse eigenmodes (${\bm A} \cdot  \hat{{\bm z}} = 0$) with spectra
\begin{equation}
\omega_\pm(k) = \sqrt{ \frac12\left[(2 + c) v^2 k^2 + \omega_\mathrm{L}^2 \pm \sqrt{c^2 v^4 k^4 + 2 (2 + c) v^2 k^2 \omega_\mathrm{L}^2 + \omega_\mathrm{L}^4}\right]}.
\end{equation}
Since $\omega_+(0) = \omega_\mathrm{L}$ and $\omega_-(0) = 0$, these modes correspond to optical magnons and acoustic magnons, respectively (cf. Fig.\ \ref{3He_modes}).
Furthermore, there is a longitudinal mode (${\bm A} \parallel \hat{{\bm z}}$) with spectrum
\begin{equation}
\omega_z(k) = \sqrt{\Omega_B^2 + (1 + c) v^2 k^2}.
\end{equation}
Since $\omega_z(0) = \Omega_B$, this corresponds to the light Higgs mode (cf. Fig.\ \ref{3He_modes}).

The above analysis helps us to understand the physics behind Eq.\ (\ref{eq:eom}) qualitatively. Since $\bm \rho \parallel \hat{{\bm z}}$, the vortex can only excite light Higgs modes. Furthermore, since $\bm \rho \propto e^{-i \omega_\mathrm{L} t}$, the vortex can only excite modes with $\omega(k) = \omega_\mathrm{L}$. Due to the fact that the minimum of $\omega_z(k)$ is $\Omega_B$, no excitations are produced when $\omega_\mathrm{L} < \Omega_B$. Conversely, when $\omega_\mathrm{L} \geq \Omega_B$, the vortex excites light Higgs modes, transferring energy away from the condensate, and thus dissipating the coherently precessing magnetisation towards the equilibrium.

\subsection*{The vortex-Higgs mechanism}
Let us then solve Eq.\ (\ref{eq:eom}) for $\bm \theta_2$. Due to the form of $\bm \rho$ [see Eq.\ (\ref{eq:rho_vector})], we can write the solution as
\begin{equation}
\bm \theta_2(\bm r, t) = \Re \left\{ e^{-i \omega_\mathrm{L} t} \tilde{\theta}_2(r, \varphi) \right\} \hat{{\bm z}},
\end{equation}
where $\tilde{\theta_2}$ satisfies the inhomogeneous Helmholtz equation
\begin{equation}
\nabla^2 \tilde{\theta}_2 + \frac{\omega_\mathrm{L}^2 - \Omega_B^2}{(1 + c) v^2} \tilde{\theta}_2 = - \frac{\tilde{\rho}}{(1 + c) v^2}.
\end{equation}
One way to solve this is to make use of the Fourier transform, as was done in Sec.\ IV in Ref.\ \onlinecite{laine2018}. As a result, we obtain
\begin{equation}
\begin{split}
\tilde{\theta}_2(r, \varphi) = -\frac{1}{\sqrt{10}} \frac{\beta C \Omega_B^2}{(1 + c)^2 v^2} \Bigg\{& \left[ (1 + c) e^{i \varphi} + \left(1 + \frac{c}{2} \right) e^{-i \varphi} \right] \left[ -\frac{1}{r K^2} + \frac{i \pi}{2 K} H_1^{(1)}(K r)  \right] \\
&+ \frac{c}{2} e^{3 i \varphi} \left[ -\frac{1}{r K^2} -\frac{8}{r^3 K^4} + \frac{i \pi}{2 K} H_3^{(1)}(K r)  \right] \Bigg\},
\end{split}
\end{equation}
where $K^2 = (\omega_\mathrm{L}^2 - \Omega_B^2) / (1 + c) v^2$ and $H_n^{(1)}(x)$ are Hankel functions of the first kind.

To calculate the energy transferred to the light Higgs mode, we integrate the energy flux density vector $\bm \Sigma$ over a cylindrical surface of radius $r$ centred at the vortex axis. Since $\bm \theta_2 \parallel \hat{{\bm z}}$, $\bm \Sigma$ takes a simple form
\begin{equation}
\bm \Sigma = - \frac{\chi v^2}{\mu_0 \gamma^2} (1 + c) \dot{\theta}_2 \bm \nabla \theta_2,
\end{equation}
where we denote $\theta_2 = \bm \theta_2 \cdot \hat{{\bm z}}$. Plugging in the solution for $\bm \theta_2$, we find that the time-averaged energy flux through the surface, per vortex length, is given by
\begin{equation}
P(r) = \frac{\chi \Omega_B^4}{\mu_0 \gamma^2} \frac{\pi^2}{40} \frac{\omega_\mathrm{L} \mathcal{H}(\omega_\mathrm{L} - \Omega_B)}{\omega_\mathrm{L}^2 - \Omega_B^2} \beta^2 C^2 \left\{\frac{3 c^2 + 6c + 4}{(1 + c)^2} -2 \frac{2 + c}{1 + c} J_0(K r) - \frac{2 c^2}{(1 + c)^2} \frac{J_1(K r)}{K r} \right\},
\end{equation}
where $\mathcal{H}(x)$ is the Heaviside step function and $J_n(x)$ are Bessel functions of the first kind. To calculate the total rate (per vortex length) at which energy is transferred to the light Higgs mode, we take the limit $P \equiv \lim_{r \to \infty} P(r)$ and find
\begin{equation}\label{eq:P}
P = \frac{\chi \Omega_B^4}{\mu_0 \gamma^2} \frac{\pi^2}{40} \frac{\omega_\mathrm{L} \mathcal{H}(\omega_\mathrm{L} - \Omega_B)}{\omega_\mathrm{L}^2 - \Omega_B^2} \beta^2 C^2 \frac{3 c^2 + 6c + 4}{(1 + c)^2}.
\end{equation}
This analysis ignores the finite dimensions of the magnon BEC in the $z$ direction and, consequently, all the power is emitted perpendicular to the vortex line. The power is maximum in the $x$ direction (perpendicular to the line connecting the vortex sub-cores, see Fig.~4 in Ref.~\onlinecite{laine2018}).

\subsection*{Vortex-Higgs relaxation time}
Let us consider what happens to the uniformly precessing magnetisation if each vortex line dissipates energy at the rate given by Eq.\ (\ref{eq:P}).

In a rotating container, the areal density of vortices is given by $n_v = 2 \Omega / \kappa$, where $\Omega$ is the angular velocity of the container, $\kappa = \pi \hbar / m_0$ the circulation quantum, and $m_0$ the mass of a $^3$He atom. One vortex thus occupies an area $A_v = 1/n_v = \kappa / 2 \Omega$. The amount of energy stored in the uniformly precessing magnetisation per vortex length is therefore given by $\mathcal{E} A_v$, where $\mathcal{E} = \chi \omega_\mathrm{L}^2 \beta_{\bm  M}^2 / 2 \mu_0 \gamma^2 +{\rm  const.}$ is the energy density. Thus, we must have $\dot{\mathcal{E}} A_v = -P$. This yields
\begin{equation}\label{exp_relax}
\dot{\beta}_{\bm  M} = - \tau^{-1} \beta_{\bm  M},
\end{equation}
where
\begin{equation}\label{final_relax}
\frac{1}{\tau} = \frac{\pi^2}{20} \Omega \frac{\Omega_B^4}{\kappa} \frac{\mathcal{H}(\omega_\mathrm{L} - \Omega_B)}{\omega_\mathrm{L} (\omega_\mathrm{L}^2 - \Omega_B^2)} \frac{3 c^2 + 6c + 4}{(1 + c)^2} C^2.
\end{equation}
We see that $\beta_{\bm  M}$ relaxes exponentially towards the equilibrium, with relaxation time $\tau$.

The relaxation rate (\ref{final_relax}) has been derived assuming a constant $\beta_{\bm M}$. It remains valid for a non-uniform distribution $\beta_{\bm M}({\bm r})$ as long as the variation of $\beta_{\bm M}$ within each vortex unit cell can be neglected. Moreover, because the radiation is effectively generated within a dipole length from the vortex axis \cite{laine2018}, it may be sufficient that $\beta_{\bm M}$ is nearly constant within this region, which at practical rotation speeds is smaller than vortex unit cell.

We see that the relaxation rate $\tau^{-1}$ (\ref{final_relax}) depends linearly on the number of vortices, as it is proportional to the  angular velocity $\Omega$ of the rotation. The relaxation rate depends on the magnetic field via the Larmor frequency $\omega_\mathrm{L}=\gamma H$. It vanishes at $\omega_\mathrm{L}<\Omega_B$ because of the energy gap of the light Higgs quasiparticles [Fig.\ 1(a)], as expressed mathematically in Eq.~(\ref{final_relax}) by the step function $\mathcal{H}(\omega_\mathrm{L} - \Omega_B)$.  For a quantitative evaluation of $\tau^{-1}$ we need  values of $\Omega_B$, $C$ and $c$, which are functions of pressure. The Leggett frequency $\Omega_B$ can be extracted from NMR experiments as discussed in Ref.\ \cite{thuneberg2001}.

The evaluation of parameters $C$ and $c$ is based on a version of the BCS theory extended to include Fermi-liquid corrections (also know as the quasiclassical theory, or the Fermi-liquid theory of superfluidity) \cite{serene1983quasiclassical}. 
The order-parameters rotation amplitude $C$ (\ref{e.asymp}) can be calculated by solving numerically the full form of the order parameter in the vortex core including all 18 real degrees of freedom at each $x$ and $y$ in the plane perpendicular to the vortex axis. At low temperatures this has been done in Ref.~\onlinecite{silaev2015} (further details of this work will be published later). This calculation includes the effect of $F_1^s$ but neglects other Fermi-liquid interaction parameters, such as $F_2^s$, $F_1^a$ and higher. It also uses the weak coupling approximation, that is, all strong-coupling effects are neglected. The result is that $C \approx 6R_0$, and the ratio $C/R_0 $ is nearly independent of pressure. Here
\begin{eqnarray}\label{e.R0}
R_0=\frac{\hbar p_F}{2\pi m_0T_c}=\left(1+\frac13F_1^s\right)\xi_0
\end{eqnarray}
is the length scale that characterises the distance between the half cores. It differs by the effect of the  Fermi liquid parameter $F_1^s$
 from the more standard coherence length $\xi_0=\hbar v_F/2\pi T_c$, which characterises the size of a half core. Here $p_F$ and $v_F$ are the Fermi momentum and Fermi velocity. Note that $C$ is approximately equal to the distance between the half cores in the double-core vortex.

The remaining parameter $c$ characterizes the anisotropy of the spin wave velocity.  In the weak coupling approximation in the limit $T/T_c\rightarrow 0$  \cite{serene1983quasiclassical}, 
\begin{equation} c=1+\frac{\frac{1}{3}F_1^{\rm a}-\frac{1}{7}F_3^{\rm a}}
{ 2(1+\frac{1}{7}F_3^{\rm a})}.
\label{e.3.6}\end{equation}
Measurements indicate $F_1^{\rm a}$ varies between -0.55 and -1.0 \cite{zavjalov2015measurements}. Assuming negligible $F_3^{\rm a}$, we find that the effect of deviation of $c$ from unity on $\tau^{-1}$ (\ref{final_relax}) is approximately one per cent.


\begin{thebibliography}{49}%
\makeatletter
\providecommand \@ifxundefined [1]{%
 \@ifx{#1\undefined}
}%
\providecommand \@ifnum [1]{%
 \ifnum #1\expandafter \@firstoftwo
 \else \expandafter \@secondoftwo
 \fi
}%
\providecommand \@ifx [1]{%
 \ifx #1\expandafter \@firstoftwo
 \else \expandafter \@secondoftwo
 \fi
}%
\providecommand \natexlab [1]{#1}%
\providecommand \enquote  [1]{``#1''}%
\providecommand \bibnamefont  [1]{#1}%
\providecommand \bibfnamefont [1]{#1}%
\providecommand \citenamefont [1]{#1}%
\providecommand \href@noop [0]{\@secondoftwo}%
\providecommand \href [0]{\begingroup \@sanitize@url \@href}%
\providecommand \@href[1]{\@@startlink{#1}\@@href}%
\providecommand \@@href[1]{\endgroup#1\@@endlink}%
\providecommand \@sanitize@url [0]{\catcode `\\12\catcode `\$12\catcode
  `\&12\catcode `\#12\catcode `\^12\catcode `\_12\catcode `\%12\relax}%
\providecommand \@@startlink[1]{}%
\providecommand \@@endlink[0]{}%
\providecommand \url  [0]{\begingroup\@sanitize@url \@url }%
\providecommand \@url [1]{\endgroup\@href {#1}{\urlprefix }}%
\providecommand \urlprefix  [0]{URL }%
\providecommand \Eprint [0]{\href }%
\providecommand \doibase [0]{http://dx.doi.org/}%
\providecommand \selectlanguage [0]{\@gobble}%
\providecommand \bibinfo  [0]{\@secondoftwo}%
\providecommand \bibfield  [0]{\@secondoftwo}%
\providecommand \translation [1]{[#1]}%
\providecommand \BibitemOpen [0]{}%
\providecommand \bibitemStop [0]{}%
\providecommand \bibitemNoStop [0]{.\EOS\space}%
\providecommand \EOS [0]{\spacefactor3000\relax}%
\providecommand \BibitemShut  [1]{\csname bibitem#1\endcsname}%
\let\auto@bib@innerbib\@empty
\bibitem [{\citenamefont {Volovik}(2003)}]{VolovikBook}%
  \BibitemOpen
  \bibfield  {author} {\bibinfo {author} {\bibfnamefont {G.~E.}\ \bibnamefont
  {Volovik}},\ }\href {\doibase 978-0-19-850782-6} {\emph {\bibinfo {title}
  {The Universe in a Helium Droplet}}}\ (\bibinfo  {publisher} {Oxford
  University Press},\ \bibinfo {year} {2003})\BibitemShut {NoStop}%
\bibitem [{\citenamefont {Bradley}\ \emph {et~al.}(2016)\citenamefont
  {Bradley}, \citenamefont {Fisher}, \citenamefont {Gu{\'e}nault},
  \citenamefont {Haley}, \citenamefont {Lawson}, \citenamefont {Pickett},
  \citenamefont {Schanen}, \citenamefont {Skyba}, \citenamefont {Tsepelin},\
  and\ \citenamefont {Zmeev}}]{bradley2016breaking}%
  \BibitemOpen
  \bibfield  {author} {\bibinfo {author} {\bibfnamefont {David~Ian}\
  \bibnamefont {Bradley}}, \bibinfo {author} {\bibfnamefont {Shaun~Neil}\
  \bibnamefont {Fisher}}, \bibinfo {author} {\bibfnamefont {Anthony~Michael}\
  \bibnamefont {Gu{\'e}nault}}, \bibinfo {author} {\bibfnamefont
  {Richard~Peter}\ \bibnamefont {Haley}}, \bibinfo {author} {\bibfnamefont
  {CR}~\bibnamefont {Lawson}}, \bibinfo {author} {\bibfnamefont
  {George~Richard}\ \bibnamefont {Pickett}}, \bibinfo {author} {\bibfnamefont
  {Roch}\ \bibnamefont {Schanen}}, \bibinfo {author} {\bibfnamefont {Maros}\
  \bibnamefont {Skyba}}, \bibinfo {author} {\bibfnamefont {Viktor}\
  \bibnamefont {Tsepelin}}, \ and\ \bibinfo {author} {\bibfnamefont
  {DE}~\bibnamefont {Zmeev}},\ }\bibfield  {title} {\enquote {\bibinfo {title}
  {Breaking the superfluid speed limit in a fermionic condensate},}\
  }\href@noop {} {\bibfield  {journal} {\bibinfo  {journal} {Nature Physics}\
  }\textbf {\bibinfo {volume} {12}},\ \bibinfo {pages} {1017--1021} (\bibinfo
  {year} {2016})}\BibitemShut {NoStop}%
\bibitem [{\citenamefont {Kuorelahti}\ \emph {et~al.}(2018)\citenamefont
  {Kuorelahti}, \citenamefont {Laine},\ and\ \citenamefont
  {Thuneberg}}]{PhysRevB.98.144512}%
  \BibitemOpen
  \bibfield  {author} {\bibinfo {author} {\bibfnamefont {J.~A.}\ \bibnamefont
  {Kuorelahti}}, \bibinfo {author} {\bibfnamefont {S.~M.}\ \bibnamefont
  {Laine}}, \ and\ \bibinfo {author} {\bibfnamefont {E.~V.}\ \bibnamefont
  {Thuneberg}},\ }\bibfield  {title} {\enquote {\bibinfo {title} {Models for
  supercritical motion in a superfluid fermi liquid},}\ }\href {\doibase
  10.1103/PhysRevB.98.144512} {\bibfield  {journal} {\bibinfo  {journal} {Phys.
  Rev. B}\ }\textbf {\bibinfo {volume} {98}},\ \bibinfo {pages} {144512}
  (\bibinfo {year} {2018})}\BibitemShut {NoStop}%
\bibitem [{\citenamefont {Autti}\ \emph
  {et~al.}(2020{\natexlab{a}})\citenamefont {Autti}, \citenamefont {M\"akinen},
  \citenamefont {Rysti}, \citenamefont {Volovik}, \citenamefont {Zavjalov},\
  and\ \citenamefont {Eltsov}}]{PhysRevResearch.2.033013}%
  \BibitemOpen
  \bibfield  {author} {\bibinfo {author} {\bibfnamefont {S.}~\bibnamefont
  {Autti}}, \bibinfo {author} {\bibfnamefont {J.~T.}\ \bibnamefont
  {M\"akinen}}, \bibinfo {author} {\bibfnamefont {J.}~\bibnamefont {Rysti}},
  \bibinfo {author} {\bibfnamefont {G.~E.}\ \bibnamefont {Volovik}}, \bibinfo
  {author} {\bibfnamefont {V.~V.}\ \bibnamefont {Zavjalov}}, \ and\ \bibinfo
  {author} {\bibfnamefont {V.~B.}\ \bibnamefont {Eltsov}},\ }\bibfield  {title}
  {\enquote {\bibinfo {title} {Exceeding the landau speed limit with
  topological bogoliubov fermi surfaces},}\ }\href {\doibase
  10.1103/PhysRevResearch.2.033013} {\bibfield  {journal} {\bibinfo  {journal}
  {Phys. Rev. Research}\ }\textbf {\bibinfo {volume} {2}},\ \bibinfo {pages}
  {033013} (\bibinfo {year} {2020}{\natexlab{a}})}\BibitemShut {NoStop}%
\bibitem [{\citenamefont {Autti}\ \emph
  {et~al.}(2020{\natexlab{b}})\citenamefont {Autti}, \citenamefont {Ahlstrom},
  \citenamefont {Haley}, \citenamefont {Jennings}, \citenamefont {Pickett},
  \citenamefont {Poole}, \citenamefont {Schanen}, \citenamefont {Soldatov},
  \citenamefont {Tsepelin}, \citenamefont {Vonka}, \citenamefont {Wilcox},
  \citenamefont {Woods},\ and\ \citenamefont {Zmeev}}]{autti2020fundamental}%
  \BibitemOpen
  \bibfield  {author} {\bibinfo {author} {\bibfnamefont {S.}~\bibnamefont
  {Autti}}, \bibinfo {author} {\bibfnamefont {S.L.}\ \bibnamefont {Ahlstrom}},
  \bibinfo {author} {\bibfnamefont {R.P.}\ \bibnamefont {Haley}}, \bibinfo
  {author} {\bibfnamefont {A.}~\bibnamefont {Jennings}}, \bibinfo {author}
  {\bibfnamefont {G.R.}\ \bibnamefont {Pickett}}, \bibinfo {author}
  {\bibfnamefont {M.}~\bibnamefont {Poole}}, \bibinfo {author} {\bibfnamefont
  {R.}~\bibnamefont {Schanen}}, \bibinfo {author} {\bibfnamefont {A.A.}\
  \bibnamefont {Soldatov}}, \bibinfo {author} {\bibfnamefont {V.}~\bibnamefont
  {Tsepelin}}, \bibinfo {author} {\bibfnamefont {J.}~\bibnamefont {Vonka}},
  \bibinfo {author} {\bibfnamefont {T.}~\bibnamefont {Wilcox}}, \bibinfo
  {author} {\bibfnamefont {A.J.}\ \bibnamefont {Woods}}, \ and\ \bibinfo
  {author} {\bibfnamefont {D.E.}\ \bibnamefont {Zmeev}},\ }\bibfield  {title}
  {\enquote {\bibinfo {title} {Fundamental dissipation due to bound fermions in
  the zero-temperature limit},}\ }\href {\doibase 10.1038/s41467-020-18499-1}
  {\bibfield  {journal} {\bibinfo  {journal} {Nature Communications}\ }\textbf
  {\bibinfo {volume} {11}} (\bibinfo {year} {2020}{\natexlab{b}}),\
  10.1038/s41467-020-18499-1}\BibitemShut {NoStop}%
\bibitem [{\citenamefont {Bunkov}\ \emph {et~al.}(1992)\citenamefont {Bunkov},
  \citenamefont {Fisher}, \citenamefont {Gu\'enault},\ and\ \citenamefont
  {Pickett}}]{1992_ppd}%
  \BibitemOpen
  \bibfield  {author} {\bibinfo {author} {\bibfnamefont {Yu.~M.}\ \bibnamefont
  {Bunkov}}, \bibinfo {author} {\bibfnamefont {S.~N.}\ \bibnamefont {Fisher}},
  \bibinfo {author} {\bibfnamefont {A.~M.}\ \bibnamefont {Gu\'enault}}, \ and\
  \bibinfo {author} {\bibfnamefont {G.~R.}\ \bibnamefont {Pickett}},\
  }\bibfield  {title} {\enquote {\bibinfo {title} {Persistent spin precession
  in {$^{3}\text{He-B}$} in the regime of vanishing quasiparticle density},}\
  }\href {\doibase 10.1103/PhysRevLett.69.3092} {\bibfield  {journal} {\bibinfo
   {journal} {Phys. Rev. Lett.}\ }\textbf {\bibinfo {volume} {69}},\ \bibinfo
  {pages} {3092--3095} (\bibinfo {year} {1992})}\BibitemShut {NoStop}%
\bibitem [{\citenamefont {Fisher}\ \emph {et~al.}(2000)\citenamefont {Fisher},
  \citenamefont {Gu\'enault}, \citenamefont {Hale}, \citenamefont {Pickett},
  \citenamefont {Reeves},\ and\ \citenamefont {Tvalashvili}}]{2000_ppd}%
  \BibitemOpen
  \bibfield  {author} {\bibinfo {author} {\bibfnamefont {S.N.}\ \bibnamefont
  {Fisher}}, \bibinfo {author} {\bibfnamefont {A.M.}\ \bibnamefont
  {Gu\'enault}}, \bibinfo {author} {\bibfnamefont {A.J.}\ \bibnamefont {Hale}},
  \bibinfo {author} {\bibfnamefont {G.R.}\ \bibnamefont {Pickett}}, \bibinfo
  {author} {\bibfnamefont {P.A.}\ \bibnamefont {Reeves}}, \ and\ \bibinfo
  {author} {\bibfnamefont {G.}~\bibnamefont {Tvalashvili}},\ }\bibfield
  {title} {\enquote {\bibinfo {title} {Thirty-minute coherence in free
  induction decay signals in superfluid {$^{3}\text{He-B}$}},}\ }\href
  {\doibase 10.1023/A:1017520806966} {\bibfield  {journal} {\bibinfo  {journal}
  {J. Low Temp. Phys.}\ }\textbf {\bibinfo {volume} {121}},\ \bibinfo {pages}
  {303--308} (\bibinfo {year} {2000})}\BibitemShut {NoStop}%
\bibitem [{\citenamefont {Autti}\ \emph {et~al.}(2012)\citenamefont {Autti},
  \citenamefont {Bunkov}, \citenamefont {Eltsov}, \citenamefont {Heikkinen},
  \citenamefont {Hosio}, \citenamefont {Hunger}, \citenamefont {Krusius},\ and\
  \citenamefont {Volovik}}]{magnon_trap_mod}%
  \BibitemOpen
  \bibfield  {author} {\bibinfo {author} {\bibfnamefont {S.}~\bibnamefont
  {Autti}}, \bibinfo {author} {\bibfnamefont {Yu.~M.}\ \bibnamefont {Bunkov}},
  \bibinfo {author} {\bibfnamefont {V.~B.}\ \bibnamefont {Eltsov}}, \bibinfo
  {author} {\bibfnamefont {P.~J.}\ \bibnamefont {Heikkinen}}, \bibinfo {author}
  {\bibfnamefont {J.~J.}\ \bibnamefont {Hosio}}, \bibinfo {author}
  {\bibfnamefont {P.}~\bibnamefont {Hunger}}, \bibinfo {author} {\bibfnamefont
  {M.}~\bibnamefont {Krusius}}, \ and\ \bibinfo {author} {\bibfnamefont
  {G.~E.}\ \bibnamefont {Volovik}},\ }\bibfield  {title} {\enquote {\bibinfo
  {title} {Self-trapping of magnon {$\text{Bose-Einstein}$} condensates in the
  ground state and on excited levels: From harmonic to box confinement},}\
  }\href {\doibase 10.1103/PhysRevLett.108.145303} {\bibfield  {journal}
  {\bibinfo  {journal} {Phys. Rev. Lett.}\ }\textbf {\bibinfo {volume} {108}},\
  \bibinfo {pages} {145303} (\bibinfo {year} {2012})}\BibitemShut {NoStop}%
\bibitem [{\citenamefont {Autti}\ \emph
  {et~al.}(2018{\natexlab{a}})\citenamefont {Autti}, \citenamefont {Dmitriev},
  \citenamefont {M\"akinen}, \citenamefont {Rysti}, \citenamefont {Soldatov},
  \citenamefont {Volovik}, \citenamefont {Yudin},\ and\ \citenamefont
  {Eltsov}}]{PhysRevLett.121.025303}%
  \BibitemOpen
  \bibfield  {author} {\bibinfo {author} {\bibfnamefont {S.}~\bibnamefont
  {Autti}}, \bibinfo {author} {\bibfnamefont {V.~V.}\ \bibnamefont {Dmitriev}},
  \bibinfo {author} {\bibfnamefont {J.~T.}\ \bibnamefont {M\"akinen}}, \bibinfo
  {author} {\bibfnamefont {J.}~\bibnamefont {Rysti}}, \bibinfo {author}
  {\bibfnamefont {A.~A.}\ \bibnamefont {Soldatov}}, \bibinfo {author}
  {\bibfnamefont {G.~E.}\ \bibnamefont {Volovik}}, \bibinfo {author}
  {\bibfnamefont {A.~N.}\ \bibnamefont {Yudin}}, \ and\ \bibinfo {author}
  {\bibfnamefont {V.~B.}\ \bibnamefont {Eltsov}},\ }\bibfield  {title}
  {\enquote {\bibinfo {title} {Bose-einstein condensation of magnons and spin
  superfluidity in the polar phase of $^{3}\mathrm{He}$},}\ }\href@noop {}
  {\bibfield  {journal} {\bibinfo  {journal} {Phys. Rev. Lett.}\ }\textbf
  {\bibinfo {volume} {121}},\ \bibinfo {pages} {025303} (\bibinfo {year}
  {2018}{\natexlab{a}})}\BibitemShut {NoStop}%
\bibitem [{\citenamefont {Murakawa}\ \emph {et~al.}(2011)\citenamefont
  {Murakawa}, \citenamefont {Wada}, \citenamefont {Tamura}, \citenamefont
  {Wasai}, \citenamefont {Saitoh}, \citenamefont {Aoki}, \citenamefont
  {Nomura}, \citenamefont {Okuda}, \citenamefont {Nagato}, \citenamefont
  {Yamamoto}, \citenamefont {Higashitani},\ and\ \citenamefont
  {Nagai}}]{Muarakawa2011}%
  \BibitemOpen
  \bibfield  {author} {\bibinfo {author} {\bibfnamefont {Satoshi}\ \bibnamefont
  {Murakawa}}, \bibinfo {author} {\bibfnamefont {Yuichiro}\ \bibnamefont
  {Wada}}, \bibinfo {author} {\bibfnamefont {Yuta}\ \bibnamefont {Tamura}},
  \bibinfo {author} {\bibfnamefont {Masahiro}\ \bibnamefont {Wasai}}, \bibinfo
  {author} {\bibfnamefont {Masamichi}\ \bibnamefont {Saitoh}}, \bibinfo
  {author} {\bibfnamefont {Yuki}\ \bibnamefont {Aoki}}, \bibinfo {author}
  {\bibfnamefont {Ryuji}\ \bibnamefont {Nomura}}, \bibinfo {author}
  {\bibfnamefont {Yuichi}\ \bibnamefont {Okuda}}, \bibinfo {author}
  {\bibfnamefont {Yasushi}\ \bibnamefont {Nagato}}, \bibinfo {author}
  {\bibfnamefont {Mikio}\ \bibnamefont {Yamamoto}}, \bibinfo {author}
  {\bibfnamefont {Seiji}\ \bibnamefont {Higashitani}}, \ and\ \bibinfo {author}
  {\bibfnamefont {Katsuhiko}\ \bibnamefont {Nagai}},\ }\bibfield  {title}
  {\enquote {\bibinfo {title} {Surface {Majorana} cone of the superfluid
  $^{3}\text{He}$ $\text{B}$ phase},}\ }\href {\doibase 10.1143/JPSJ.80.013602}
  {\bibfield  {journal} {\bibinfo  {journal} {J. Phys. Soc. Jpn.}\ }\textbf
  {\bibinfo {volume} {80}},\ \bibinfo {pages} {013602} (\bibinfo {year}
  {2011})}\BibitemShut {NoStop}%
\bibitem [{\citenamefont {Rosenstein}\ \emph {et~al.}(2013)\citenamefont
  {Rosenstein}, \citenamefont {Shapiro},\ and\ \citenamefont
  {Shapiro}}]{Shapiro2013}%
  \BibitemOpen
  \bibfield  {author} {\bibinfo {author} {\bibfnamefont {B}~\bibnamefont
  {Rosenstein}}, \bibinfo {author} {\bibfnamefont {I}~\bibnamefont {Shapiro}},
  \ and\ \bibinfo {author} {\bibfnamefont {B~Ya}\ \bibnamefont {Shapiro}},\
  }\bibfield  {title} {\enquote {\bibinfo {title} {Effect of nanoholes on the
  vortex core fermion spectrum and heat transport in p-wave superconductors},}\
  }\href@noop {} {\bibfield  {journal} {\bibinfo  {journal} {J. Phys.: Condens.
  Matter}\ }\textbf {\bibinfo {volume} {25}},\ \bibinfo {pages} {075701}
  (\bibinfo {year} {2013})}\BibitemShut {NoStop}%
\bibitem [{\citenamefont {Eltsov}\ and\ \citenamefont
  {L’vov}(2020)}]{eltsov2020amplitude}%
  \BibitemOpen
  \bibfield  {author} {\bibinfo {author} {\bibfnamefont {Vladimir~B}\
  \bibnamefont {Eltsov}}\ and\ \bibinfo {author} {\bibfnamefont
  {VS}~\bibnamefont {L’vov}},\ }\bibfield  {title} {\enquote {\bibinfo
  {title} {Amplitude of waves in the {Kelvin}-wave cascade},}\ }\href@noop {}
  {\bibfield  {journal} {\bibinfo  {journal} {JETP Letters}\ ,\ \bibinfo
  {pages} {1--3}} (\bibinfo {year} {2020})}\BibitemShut {NoStop}%
\bibitem [{\citenamefont {L’vov}\ and\ \citenamefont
  {Nazarenko}(2010)}]{l2010spectrum}%
  \BibitemOpen
  \bibfield  {author} {\bibinfo {author} {\bibfnamefont {Victor~S}\
  \bibnamefont {L’vov}}\ and\ \bibinfo {author} {\bibfnamefont {Sergey}\
  \bibnamefont {Nazarenko}},\ }\bibfield  {title} {\enquote {\bibinfo {title}
  {Spectrum of {Kelvin}-wave turbulence in superfluids},}\ }\href@noop {}
  {\bibfield  {journal} {\bibinfo  {journal} {JETP Letters}\ }\textbf {\bibinfo
  {volume} {91}},\ \bibinfo {pages} {428--434} (\bibinfo {year}
  {2010})}\BibitemShut {NoStop}%
\bibitem [{\citenamefont {Autti}\ \emph
  {et~al.}(2018{\natexlab{b}})\citenamefont {Autti}, \citenamefont {Eltsov},\
  and\ \citenamefont {Volovik}}]{PhysRevLett.120.215301}%
  \BibitemOpen
  \bibfield  {author} {\bibinfo {author} {\bibfnamefont {S.}~\bibnamefont
  {Autti}}, \bibinfo {author} {\bibfnamefont {V.~B.}\ \bibnamefont {Eltsov}}, \
  and\ \bibinfo {author} {\bibfnamefont {G.~E.}\ \bibnamefont {Volovik}},\
  }\bibfield  {title} {\enquote {\bibinfo {title} {Observation of a time
  quasicrystal and its transition to a superfluid time crystal},}\ }\href
  {\doibase 10.1103/PhysRevLett.120.215301} {\bibfield  {journal} {\bibinfo
  {journal} {Phys. Rev. Lett.}\ }\textbf {\bibinfo {volume} {120}},\ \bibinfo
  {pages} {215301} (\bibinfo {year} {2018}{\natexlab{b}})}\BibitemShut
  {NoStop}%
\bibitem [{\citenamefont {Kreil}\ \emph
  {et~al.}(2019{\natexlab{a}})\citenamefont {Kreil}, \citenamefont
  {Musiienko-Shmarova}, \citenamefont {Eggert}, \citenamefont {Serga},
  \citenamefont {Hillebrands}, \citenamefont {Bozhko}, \citenamefont
  {Pomyalov},\ and\ \citenamefont {L'vov}}]{PhysRevB.100.020406}%
  \BibitemOpen
  \bibfield  {author} {\bibinfo {author} {\bibfnamefont {Alexander J.~E.}\
  \bibnamefont {Kreil}}, \bibinfo {author} {\bibfnamefont {Halyna~Yu.}\
  \bibnamefont {Musiienko-Shmarova}}, \bibinfo {author} {\bibfnamefont
  {Sebastian}\ \bibnamefont {Eggert}}, \bibinfo {author} {\bibfnamefont
  {Alexander~A.}\ \bibnamefont {Serga}}, \bibinfo {author} {\bibfnamefont
  {Burkard}\ \bibnamefont {Hillebrands}}, \bibinfo {author} {\bibfnamefont
  {Dmytro~A.}\ \bibnamefont {Bozhko}}, \bibinfo {author} {\bibfnamefont {Anna}\
  \bibnamefont {Pomyalov}}, \ and\ \bibinfo {author} {\bibfnamefont
  {Victor~S.}\ \bibnamefont {L'vov}},\ }\bibfield  {title} {\enquote {\bibinfo
  {title} {Tunable space-time crystal in room-temperature
  magnetodielectrics},}\ }\href {\doibase 10.1103/PhysRevB.100.020406}
  {\bibfield  {journal} {\bibinfo  {journal} {Phys. Rev. B}\ }\textbf {\bibinfo
  {volume} {100}},\ \bibinfo {pages} {020406} (\bibinfo {year}
  {2019}{\natexlab{a}})}\BibitemShut {NoStop}%
\bibitem [{\citenamefont {Autti}\ \emph
  {et~al.}(2020{\natexlab{c}})\citenamefont {Autti}, \citenamefont {Heikkinen},
  \citenamefont {Mäkinen}, \citenamefont {Volovik}, \citenamefont {Zavjalov},\
  and\ \citenamefont {Eltsov}}]{autti2020ac}%
  \BibitemOpen
  \bibfield  {author} {\bibinfo {author} {\bibfnamefont {S.}~\bibnamefont
  {Autti}}, \bibinfo {author} {\bibfnamefont {P.~J.}\ \bibnamefont
  {Heikkinen}}, \bibinfo {author} {\bibfnamefont {J.~T.}\ \bibnamefont
  {Mäkinen}}, \bibinfo {author} {\bibfnamefont {G.~E.}\ \bibnamefont
  {Volovik}}, \bibinfo {author} {\bibfnamefont {V.~V.}\ \bibnamefont
  {Zavjalov}}, \ and\ \bibinfo {author} {\bibfnamefont {V.~B.}\ \bibnamefont
  {Eltsov}},\ }\bibfield  {title} {\enquote {\bibinfo {title} {{AC} {Josephson}
  effect between two superfluid time crystals},}\ }\href {\doibase
  10.1038/s41563-020-0780-y} {\bibfield  {journal} {\bibinfo  {journal} {Nature
  Materials}\ } (\bibinfo {year} {2020}{\natexlab{c}}),\
  10.1038/s41563-020-0780-y}\BibitemShut {NoStop}%
\bibitem [{\citenamefont {Leggett}(1975)}]{leggett1975theoretical}%
  \BibitemOpen
  \bibfield  {author} {\bibinfo {author} {\bibfnamefont {Anthony~J}\
  \bibnamefont {Leggett}},\ }\bibfield  {title} {\enquote {\bibinfo {title} {A
  theoretical description of the new phases of liquid he 3},}\ }\href@noop {}
  {\bibfield  {journal} {\bibinfo  {journal} {Reviews of Modern Physics}\
  }\textbf {\bibinfo {volume} {47}},\ \bibinfo {pages} {331} (\bibinfo {year}
  {1975})}\BibitemShut {NoStop}%
\bibitem [{\citenamefont {Vollhardt}\ and\ \citenamefont
  {W{\"o}lfle}(2013)}]{vollhardt1990}%
  \BibitemOpen
  \bibfield  {author} {\bibinfo {author} {\bibfnamefont {D.}~\bibnamefont
  {Vollhardt}}\ and\ \bibinfo {author} {\bibfnamefont {P.}~\bibnamefont
  {W{\"o}lfle}},\ }\href {https://books.google.fi/books?id=4grCAgAAQBAJ} {\emph
  {\bibinfo {title} {The Superfluid Phases of Helium 3}}}\ (\bibinfo
  {publisher} {Dover Publications},\ \bibinfo {year} {2013})\BibitemShut
  {NoStop}%
\bibitem [{\citenamefont {Zavjalov}\ \emph {et~al.}(2016)\citenamefont
  {Zavjalov}, \citenamefont {Autti}, \citenamefont {Eltsov}, \citenamefont
  {Heikkinen},\ and\ \citenamefont {Volovik}}]{HiggsNComm}%
  \BibitemOpen
  \bibfield  {author} {\bibinfo {author} {\bibfnamefont {V.~V.}\ \bibnamefont
  {Zavjalov}}, \bibinfo {author} {\bibfnamefont {S.}~\bibnamefont {Autti}},
  \bibinfo {author} {\bibfnamefont {V.~B.}\ \bibnamefont {Eltsov}}, \bibinfo
  {author} {\bibfnamefont {P.~J.}\ \bibnamefont {Heikkinen}}, \ and\ \bibinfo
  {author} {\bibfnamefont {G.~E.}\ \bibnamefont {Volovik}},\ }\bibfield
  {title} {\enquote {\bibinfo {title} {Light {Higgs} channel of the resonant
  decay of magnon condensate in superfluid $^3 \text{He}$.}}\ }\href@noop {}
  {\bibfield  {journal} {\bibinfo  {journal} {Nat. Commun.}\ }\textbf {\bibinfo
  {volume} {7}},\ \bibinfo {pages} {10294} (\bibinfo {year}
  {2016})}\BibitemShut {NoStop}%
\bibitem [{\citenamefont {M\"akinen}\ and\ \citenamefont
  {Eltsov}(2018)}]{PhysRevB.97.014527}%
  \BibitemOpen
  \bibfield  {author} {\bibinfo {author} {\bibfnamefont {J.~T.}\ \bibnamefont
  {M\"akinen}}\ and\ \bibinfo {author} {\bibfnamefont {V.~B.}\ \bibnamefont
  {Eltsov}},\ }\bibfield  {title} {\enquote {\bibinfo {title} {Mutual friction
  in superfluid $^{3}\mathrm{He}\ensuremath{-}\mathrm{B}$ in the
  low-temperature regime},}\ }\href {\doibase 10.1103/PhysRevB.97.014527}
  {\bibfield  {journal} {\bibinfo  {journal} {Phys. Rev. B}\ }\textbf {\bibinfo
  {volume} {97}},\ \bibinfo {pages} {014527} (\bibinfo {year}
  {2018})}\BibitemShut {NoStop}%
\bibitem [{\citenamefont {Salomaa}\ and\ \citenamefont
  {Volovik}(1987)}]{RevModPhys.59.533}%
  \BibitemOpen
  \bibfield  {author} {\bibinfo {author} {\bibfnamefont {M.~M.}\ \bibnamefont
  {Salomaa}}\ and\ \bibinfo {author} {\bibfnamefont {G.~E.}\ \bibnamefont
  {Volovik}},\ }\bibfield  {title} {\enquote {\bibinfo {title} {Quantized
  vortices in superfluid $^{3}\mathrm{He}$},}\ }\href {\doibase
  10.1103/RevModPhys.59.533} {\bibfield  {journal} {\bibinfo  {journal} {Rev.
  Mod. Phys.}\ }\textbf {\bibinfo {volume} {59}},\ \bibinfo {pages} {533--613}
  (\bibinfo {year} {1987})}\BibitemShut {NoStop}%
\bibitem [{\citenamefont {Lounasmaa}\ and\ \citenamefont
  {Thuneberg}(1999)}]{lounasmaa1999vortices}%
  \BibitemOpen
  \bibfield  {author} {\bibinfo {author} {\bibfnamefont {Olli~V}\ \bibnamefont
  {Lounasmaa}}\ and\ \bibinfo {author} {\bibfnamefont {Erkki}\ \bibnamefont
  {Thuneberg}},\ }\bibfield  {title} {\enquote {\bibinfo {title} {Vortices in
  rotating superfluid $^3${He}},}\ }\href@noop {} {\bibfield  {journal}
  {\bibinfo  {journal} {Proceedings of the National Academy of Sciences}\
  }\textbf {\bibinfo {volume} {96}},\ \bibinfo {pages} {7760--7767} (\bibinfo
  {year} {1999})}\BibitemShut {NoStop}%
\bibitem [{\citenamefont {Thuneberg}(1987)}]{thuneberg1987}%
  \BibitemOpen
  \bibfield  {author} {\bibinfo {author} {\bibfnamefont {E.~V.}\ \bibnamefont
  {Thuneberg}},\ }\bibfield  {title} {\enquote {\bibinfo {title}
  {Ginzburg-{Landau} theory of vortices in superfluid $^{3}${He-B}},}\ }\href
  {\doibase 10.1103/PhysRevB.36.3583} {\bibfield  {journal} {\bibinfo
  {journal} {Phys. Rev. B}\ }\textbf {\bibinfo {volume} {36}},\ \bibinfo
  {pages} {3583--3597} (\bibinfo {year} {1987})}\BibitemShut {NoStop}%
\bibitem [{\citenamefont {Volovik}(1990)}]{volovik1990half}%
  \BibitemOpen
  \bibfield  {author} {\bibinfo {author} {\bibfnamefont {G}~\bibnamefont
  {Volovik}},\ }\bibfield  {title} {\enquote {\bibinfo {title} {Half-quantum
  vortices in superfluid $^3${He}-{B}},}\ }\href@noop {} {\bibfield  {journal}
  {\bibinfo  {journal} {JETP Lett}\ }\textbf {\bibinfo {volume} {52}} (\bibinfo
  {year} {1990})}\BibitemShut {NoStop}%
\bibitem [{\citenamefont {Fogelström}\ and\ \citenamefont
  {Kurkijärvi}(1995)}]{fogelstrom1995}%
  \BibitemOpen
  \bibfield  {author} {\bibinfo {author} {\bibfnamefont {M.}~\bibnamefont
  {Fogelström}}\ and\ \bibinfo {author} {\bibfnamefont {J.}~\bibnamefont
  {Kurkijärvi}},\ }\bibfield  {title} {\enquote {\bibinfo {title}
  {Quasiclassical theory of vortices in $^3${He-B}},}\ }\href {\doibase
  10.1007/BF00753614} {\bibfield  {journal} {\bibinfo  {journal} {Journal of
  Low Temperature Physics}\ }\textbf {\bibinfo {volume} {98}},\ \bibinfo
  {pages} {195--226} (\bibinfo {year} {1995})}\BibitemShut {NoStop}%
\bibitem [{\citenamefont {Silaev}\ \emph {et~al.}(2015)\citenamefont {Silaev},
  \citenamefont {Thuneberg},\ and\ \citenamefont {Fogelstr\"om}}]{silaev2015}%
  \BibitemOpen
  \bibfield  {author} {\bibinfo {author} {\bibfnamefont {M.~A.}\ \bibnamefont
  {Silaev}}, \bibinfo {author} {\bibfnamefont {E.~V.}\ \bibnamefont
  {Thuneberg}}, \ and\ \bibinfo {author} {\bibfnamefont {M.}~\bibnamefont
  {Fogelstr\"om}},\ }\bibfield  {title} {\enquote {\bibinfo {title} {Lifshitz
  transition in the double-core vortex in
  $^{3}\mathrm{He}\text{\ensuremath{-}}\mathrm{B}$},}\ }\href {\doibase
  10.1103/PhysRevLett.115.235301} {\bibfield  {journal} {\bibinfo  {journal}
  {Phys. Rev. Lett.}\ }\textbf {\bibinfo {volume} {115}},\ \bibinfo {pages}
  {235301} (\bibinfo {year} {2015})}\BibitemShut {NoStop}%
\bibitem [{\citenamefont {Kasamatsu}\ \emph {et~al.}(2019)\citenamefont
  {Kasamatsu}, \citenamefont {Mizuno}, \citenamefont {Ohmi},\ and\
  \citenamefont {Nakahara}}]{kasamatsu2019effects}%
  \BibitemOpen
  \bibfield  {author} {\bibinfo {author} {\bibfnamefont {Kenichi}\ \bibnamefont
  {Kasamatsu}}, \bibinfo {author} {\bibfnamefont {Ryota}\ \bibnamefont
  {Mizuno}}, \bibinfo {author} {\bibfnamefont {Tetsuo}\ \bibnamefont {Ohmi}}, \
  and\ \bibinfo {author} {\bibfnamefont {Mikio}\ \bibnamefont {Nakahara}},\
  }\bibfield  {title} {\enquote {\bibinfo {title} {Effects of a magnetic field
  on vortex states in superfluid $^3${He}-{B}},}\ }\href@noop {} {\bibfield
  {journal} {\bibinfo  {journal} {Physical Review B}\ }\textbf {\bibinfo
  {volume} {99}},\ \bibinfo {pages} {104513} (\bibinfo {year}
  {2019})}\BibitemShut {NoStop}%
\bibitem [{\citenamefont {Nagamura}\ and\ \citenamefont
  {Ikeda}(2019)}]{nagamura2019doublecore}%
  \BibitemOpen
  \bibfield  {author} {\bibinfo {author} {\bibfnamefont {Natsuo}\ \bibnamefont
  {Nagamura}}\ and\ \bibinfo {author} {\bibfnamefont {Ryusuke}\ \bibnamefont
  {Ikeda}},\ }\href@noop {} {\enquote {\bibinfo {title} {Double-core vortex
  stabilized by disorder in superfluid $^3${He} {B} phase in globally isotropic
  aerogel},}\ } (\bibinfo {year} {2019}),\ \Eprint
  {http://arxiv.org/abs/1905.02569} {arXiv:1905.02569 [cond-mat.supr-con]}
  \BibitemShut {NoStop}%
\bibitem [{\citenamefont {Regan}\ \emph {et~al.}(2020)\citenamefont {Regan},
  \citenamefont {Wiman},\ and\ \citenamefont {Sauls}}]{regan2020vortex}%
  \BibitemOpen
  \bibfield  {author} {\bibinfo {author} {\bibfnamefont {Robert~C}\
  \bibnamefont {Regan}}, \bibinfo {author} {\bibfnamefont {JJ}~\bibnamefont
  {Wiman}}, \ and\ \bibinfo {author} {\bibfnamefont {JA}~\bibnamefont
  {Sauls}},\ }\bibfield  {title} {\enquote {\bibinfo {title} {Vortex phase
  diagram of rotating superfluid $^3${He}-{B}},}\ }\href@noop {} {\bibfield
  {journal} {\bibinfo  {journal} {Physical Review B}\ }\textbf {\bibinfo
  {volume} {101}},\ \bibinfo {pages} {024517} (\bibinfo {year}
  {2020})}\BibitemShut {NoStop}%
\bibitem [{\citenamefont {Serene}\ and\ \citenamefont
  {Rainer}(1983)}]{serene1983quasiclassical}%
  \BibitemOpen
  \bibfield  {author} {\bibinfo {author} {\bibfnamefont {Joseph~W}\
  \bibnamefont {Serene}}\ and\ \bibinfo {author} {\bibfnamefont {Dierk}\
  \bibnamefont {Rainer}},\ }\bibfield  {title} {\enquote {\bibinfo {title} {The
  quasiclassical approach to superfluid $^3${He}},}\ }\href@noop {} {\bibfield
  {journal} {\bibinfo  {journal} {Physics Reports}\ }\textbf {\bibinfo {volume}
  {101}},\ \bibinfo {pages} {221--311} (\bibinfo {year} {1983})}\BibitemShut
  {NoStop}%
\bibitem [{\citenamefont {Bunkov}\ and\ \citenamefont
  {Volovik}(2013)}]{magnon_BEC_review}%
  \BibitemOpen
  \bibfield  {author} {\bibinfo {author} {\bibfnamefont {Yu.~M.}\ \bibnamefont
  {Bunkov}}\ and\ \bibinfo {author} {\bibfnamefont {G.~E.}\ \bibnamefont
  {Volovik}},\ }\href@noop {} {\emph {\bibinfo {title} {Novel Superfluids}}},\
  Vol.~\bibinfo {volume} {1}\ (\bibinfo  {publisher} {Oxford University Press,
  Oxford},\ \bibinfo {year} {2013})\ pp.\ \bibinfo {pages}
  {253--311}\BibitemShut {NoStop}%
\bibitem [{\citenamefont {Heikkinen}\ \emph
  {et~al.}(2014{\natexlab{a}})\citenamefont {Heikkinen}, \citenamefont {Autti},
  \citenamefont {Eltsov}, \citenamefont {Hosio}, \citenamefont {Krusius},\ and\
  \citenamefont {Zavjalov}}]{magnon_relax}%
  \BibitemOpen
  \bibfield  {author} {\bibinfo {author} {\bibfnamefont {P.~J.}\ \bibnamefont
  {Heikkinen}}, \bibinfo {author} {\bibfnamefont {S.}~\bibnamefont {Autti}},
  \bibinfo {author} {\bibfnamefont {V.B.}\ \bibnamefont {Eltsov}}, \bibinfo
  {author} {\bibfnamefont {J.J.}\ \bibnamefont {Hosio}}, \bibinfo {author}
  {\bibfnamefont {M.}~\bibnamefont {Krusius}}, \ and\ \bibinfo {author}
  {\bibfnamefont {V.V.}\ \bibnamefont {Zavjalov}},\ }\bibfield  {title}
  {\enquote {\bibinfo {title} {Relaxation of {$\text{Bose-Einstein}$}
  condensates of magnons in magneto-textural traps in superfluid
  {$^{3}\text{He-B}$}},}\ }\href {\doibase 10.1007/s10909-013-0946-y}
  {\bibfield  {journal} {\bibinfo  {journal} {J. Low Temp. Phys.}\ }\textbf
  {\bibinfo {volume} {175}},\ \bibinfo {pages} {3--16} (\bibinfo {year}
  {2014}{\natexlab{a}})}\BibitemShut {NoStop}%
\bibitem [{\citenamefont {Zavjalov}\ \emph {et~al.}(2015)\citenamefont
  {Zavjalov}, \citenamefont {Autti}, \citenamefont {Eltsov},\ and\
  \citenamefont {Heikkinen}}]{zavjalov2015measurements}%
  \BibitemOpen
  \bibfield  {author} {\bibinfo {author} {\bibfnamefont {V.V.}\ \bibnamefont
  {Zavjalov}}, \bibinfo {author} {\bibfnamefont {S.}~\bibnamefont {Autti}},
  \bibinfo {author} {\bibfnamefont {V.B.}\ \bibnamefont {Eltsov}}, \ and\
  \bibinfo {author} {\bibfnamefont {P.J.}\ \bibnamefont {Heikkinen}},\
  }\bibfield  {title} {\enquote {\bibinfo {title} {Measurements of the
  anisotropic mass of magnons confined in a harmonic trap in superfluid
  $^3${He}-{B}},}\ }\href@noop {} {\bibfield  {journal} {\bibinfo  {journal}
  {JETP letters}\ }\textbf {\bibinfo {volume} {101}},\ \bibinfo {pages}
  {802--807} (\bibinfo {year} {2015})}\BibitemShut {NoStop}%
\bibitem [{\citenamefont {Blaauwgeers}\ \emph {et~al.}(2007)\citenamefont
  {Blaauwgeers}, \citenamefont {Bla\v{z}kov\'{a}}, \citenamefont
  {\v{C}love\v{c}ko}, \citenamefont {Eltsov}, \citenamefont {de~Graaf},
  \citenamefont {Hosio}, \citenamefont {Krusius}, \citenamefont {Schmoranzer},
  \citenamefont {Schoepe}, \citenamefont {Skrbek}, \citenamefont {Skyba},
  \citenamefont {Solntsev},\ and\ \citenamefont {Zmeev}}]{2007_forks}%
  \BibitemOpen
  \bibfield  {author} {\bibinfo {author} {\bibfnamefont {R.}~\bibnamefont
  {Blaauwgeers}}, \bibinfo {author} {\bibfnamefont {M.}~\bibnamefont
  {Bla\v{z}kov\'{a}}}, \bibinfo {author} {\bibfnamefont {M.}~\bibnamefont
  {\v{C}love\v{c}ko}}, \bibinfo {author} {\bibfnamefont {V.~B.}\ \bibnamefont
  {Eltsov}}, \bibinfo {author} {\bibfnamefont {R.}~\bibnamefont {de~Graaf}},
  \bibinfo {author} {\bibfnamefont {J.}~\bibnamefont {Hosio}}, \bibinfo
  {author} {\bibfnamefont {M.}~\bibnamefont {Krusius}}, \bibinfo {author}
  {\bibfnamefont {D.}~\bibnamefont {Schmoranzer}}, \bibinfo {author}
  {\bibfnamefont {W.}~\bibnamefont {Schoepe}}, \bibinfo {author} {\bibfnamefont
  {L.}~\bibnamefont {Skrbek}}, \bibinfo {author} {\bibfnamefont
  {P.}~\bibnamefont {Skyba}}, \bibinfo {author} {\bibfnamefont {R.~E.}\
  \bibnamefont {Solntsev}}, \ and\ \bibinfo {author} {\bibfnamefont {D.~E.}\
  \bibnamefont {Zmeev}},\ }\bibfield  {title} {\enquote {\bibinfo {title}
  {Quartz tuning fork: Thermometer, pressure- and viscometer for helium
  liquids},}\ }\href {\doibase 10.1007/s10909-006-9279-4} {\bibfield  {journal}
  {\bibinfo  {journal} {J. Low Temp. Phys.}\ }\textbf {\bibinfo {volume}
  {146}},\ \bibinfo {pages} {537--562} (\bibinfo {year} {2007})}\BibitemShut
  {NoStop}%
\bibitem [{\citenamefont {Heikkinen}\ \emph
  {et~al.}(2014{\natexlab{b}})\citenamefont {Heikkinen}, \citenamefont {Autti},
  \citenamefont {Eltsov}, \citenamefont {Haley},\ and\ \citenamefont
  {Zavjalov}}]{Heikkinen2014}%
  \BibitemOpen
  \bibfield  {author} {\bibinfo {author} {\bibfnamefont {P.~J.}\ \bibnamefont
  {Heikkinen}}, \bibinfo {author} {\bibfnamefont {S.}~\bibnamefont {Autti}},
  \bibinfo {author} {\bibfnamefont {V.~B.}\ \bibnamefont {Eltsov}}, \bibinfo
  {author} {\bibfnamefont {R.~P.}\ \bibnamefont {Haley}}, \ and\ \bibinfo
  {author} {\bibfnamefont {V.~V.}\ \bibnamefont {Zavjalov}},\ }\bibfield
  {title} {\enquote {\bibinfo {title} {Microkelvin thermometry with
  $\text{Bose-Einstein}$ condensates of magnons and applications to studies of
  the $\text{AB}$ interface in superfluid $^{3}\text{He}$},}\ }\href {\doibase
  10.1007/s10909-014-1173-x} {\bibfield  {journal} {\bibinfo  {journal} {J. Low
  Temp. Phys.}\ }\textbf {\bibinfo {volume} {175}},\ \bibinfo {pages}
  {681--705} (\bibinfo {year} {2014}{\natexlab{b}})}\BibitemShut {NoStop}%
\bibitem [{\citenamefont {Fisher}\ \emph {et~al.}(2012)\citenamefont {Fisher},
  \citenamefont {Pickett}, \citenamefont {Skyba},\ and\ \citenamefont
  {Suramlishvili}}]{PhysRevB.86.024506}%
  \BibitemOpen
  \bibfield  {author} {\bibinfo {author} {\bibfnamefont {S.~N.}\ \bibnamefont
  {Fisher}}, \bibinfo {author} {\bibfnamefont {G.~R.}\ \bibnamefont {Pickett}},
  \bibinfo {author} {\bibfnamefont {P.}~\bibnamefont {Skyba}}, \ and\ \bibinfo
  {author} {\bibfnamefont {N.}~\bibnamefont {Suramlishvili}},\ }\bibfield
  {title} {\enquote {\bibinfo {title} {Decay of persistent precessing domains
  in $^{3}${He}-{B} at very low temperatures},}\ }\href {\doibase
  10.1103/PhysRevB.86.024506} {\bibfield  {journal} {\bibinfo  {journal} {Phys.
  Rev. B}\ }\textbf {\bibinfo {volume} {86}},\ \bibinfo {pages} {024506}
  (\bibinfo {year} {2012})}\BibitemShut {NoStop}%
\bibitem [{\citenamefont {Laine}\ and\ \citenamefont
  {Thuneberg}(2018)}]{laine2018}%
  \BibitemOpen
  \bibfield  {author} {\bibinfo {author} {\bibfnamefont {S.~M.}\ \bibnamefont
  {Laine}}\ and\ \bibinfo {author} {\bibfnamefont {E.~V.}\ \bibnamefont
  {Thuneberg}},\ }\bibfield  {title} {\enquote {\bibinfo {title} {Spin-wave
  radiation from vortices in
  ${}^{3}\mathrm{He}\text{\ensuremath{-}}\mathrm{B}$},}\ }\href {\doibase
  10.1103/PhysRevB.98.174516} {\bibfield  {journal} {\bibinfo  {journal} {Phys.
  Rev. B}\ }\textbf {\bibinfo {volume} {98}},\ \bibinfo {pages} {174516}
  (\bibinfo {year} {2018})}\BibitemShut {NoStop}%
\bibitem [{\citenamefont {Eltsov}\ \emph {et~al.}(2011)\citenamefont {Eltsov},
  \citenamefont {De~Graaf}, \citenamefont {Krusius},\ and\ \citenamefont
  {Zmeev}}]{eltsov2011vortex}%
  \BibitemOpen
  \bibfield  {author} {\bibinfo {author} {\bibfnamefont {V.B.}\ \bibnamefont
  {Eltsov}}, \bibinfo {author} {\bibfnamefont {R.}~\bibnamefont {De~Graaf}},
  \bibinfo {author} {\bibfnamefont {M.}~\bibnamefont {Krusius}}, \ and\
  \bibinfo {author} {\bibfnamefont {D.E.}\ \bibnamefont {Zmeev}},\ }\bibfield
  {title} {\enquote {\bibinfo {title} {Vortex core contribution to textural
  energy in $^3${He}--{B} below 0.4{T}$_\mathrm{c}$},}\ }\href@noop {}
  {\bibfield  {journal} {\bibinfo  {journal} {Journal of Low Temperature
  Physics}\ }\textbf {\bibinfo {volume} {162}},\ \bibinfo {pages} {212--225}
  (\bibinfo {year} {2011})}\BibitemShut {NoStop}%
\bibitem [{\citenamefont {Thuneberg}(2001)}]{thuneberg2001}%
  \BibitemOpen
  \bibfield  {author} {\bibinfo {author} {\bibfnamefont {E.~V.}\ \bibnamefont
  {Thuneberg}},\ }\bibfield  {title} {\enquote {\bibinfo {title} {Hydrostatic
  theory of superfluid {$^{3}\text{He-B}$}},}\ }\href {\doibase
  10.1023/A:1004898420870} {\bibfield  {journal} {\bibinfo  {journal} {J. Low
  Temp. Phys.}\ }\textbf {\bibinfo {volume} {122}},\ \bibinfo {pages}
  {657--682} (\bibinfo {year} {2001})}\BibitemShut {NoStop}%
\bibitem [{\citenamefont {Silaev}(2012)}]{PhysRevLett.108.045303}%
  \BibitemOpen
  \bibfield  {author} {\bibinfo {author} {\bibfnamefont {Mihail~A.}\
  \bibnamefont {Silaev}},\ }\bibfield  {title} {\enquote {\bibinfo {title}
  {Universal mechanism of dissipation in fermi superfluids at ultralow
  temperatures},}\ }\href {\doibase 10.1103/PhysRevLett.108.045303} {\bibfield
  {journal} {\bibinfo  {journal} {Phys. Rev. Lett.}\ }\textbf {\bibinfo
  {volume} {108}},\ \bibinfo {pages} {045303} (\bibinfo {year}
  {2012})}\BibitemShut {NoStop}%
\bibitem [{\citenamefont {Kopnin}\ and\ \citenamefont
  {Salomaa}(1991)}]{PhysRevB.44.9667}%
  \BibitemOpen
  \bibfield  {author} {\bibinfo {author} {\bibfnamefont {N.~B.}\ \bibnamefont
  {Kopnin}}\ and\ \bibinfo {author} {\bibfnamefont {M.~M.}\ \bibnamefont
  {Salomaa}},\ }\bibfield  {title} {\enquote {\bibinfo {title} {Mutual friction
  in superfluid $^{3}\mathrm{He}$: Effects of bound states in the vortex
  core},}\ }\href {\doibase 10.1103/PhysRevB.44.9667} {\bibfield  {journal}
  {\bibinfo  {journal} {Phys. Rev. B}\ }\textbf {\bibinfo {volume} {44}},\
  \bibinfo {pages} {9667--9677} (\bibinfo {year} {1991})}\BibitemShut {NoStop}%
\bibitem [{\citenamefont {Kopnin}\ and\ \citenamefont
  {Volovik}(1998)}]{kopnin1998rotating}%
  \BibitemOpen
  \bibfield  {author} {\bibinfo {author} {\bibfnamefont {NB}~\bibnamefont
  {Kopnin}}\ and\ \bibinfo {author} {\bibfnamefont {GE}~\bibnamefont
  {Volovik}},\ }\bibfield  {title} {\enquote {\bibinfo {title} {Rotating vortex
  core: An instrument for detecting core excitations},}\ }\href@noop {}
  {\bibfield  {journal} {\bibinfo  {journal} {Physical Review B}\ }\textbf
  {\bibinfo {volume} {57}},\ \bibinfo {pages} {8526} (\bibinfo {year}
  {1998})}\BibitemShut {NoStop}%
\bibitem [{\citenamefont {Kozik}\ and\ \citenamefont
  {Svistunov}(2004)}]{kozik2004kelvin}%
  \BibitemOpen
  \bibfield  {author} {\bibinfo {author} {\bibfnamefont {Evgeny}\ \bibnamefont
  {Kozik}}\ and\ \bibinfo {author} {\bibfnamefont {Boris}\ \bibnamefont
  {Svistunov}},\ }\bibfield  {title} {\enquote {\bibinfo {title} {Kelvin-wave
  cascade and decay of superfluid turbulence},}\ }\href@noop {} {\bibfield
  {journal} {\bibinfo  {journal} {Physical review letters}\ }\textbf {\bibinfo
  {volume} {92}},\ \bibinfo {pages} {035301} (\bibinfo {year}
  {2004})}\BibitemShut {NoStop}%
\bibitem [{\citenamefont {Kivotides}\ \emph {et~al.}(2001)\citenamefont
  {Kivotides}, \citenamefont {Vassilicos}, \citenamefont {Samuels},\ and\
  \citenamefont {Barenghi}}]{PhysRevLett.86.3080}%
  \BibitemOpen
  \bibfield  {author} {\bibinfo {author} {\bibfnamefont {D.}~\bibnamefont
  {Kivotides}}, \bibinfo {author} {\bibfnamefont {J.~C.}\ \bibnamefont
  {Vassilicos}}, \bibinfo {author} {\bibfnamefont {D.~C.}\ \bibnamefont
  {Samuels}}, \ and\ \bibinfo {author} {\bibfnamefont {C.~F.}\ \bibnamefont
  {Barenghi}},\ }\bibfield  {title} {\enquote {\bibinfo {title} {Kelvin waves
  cascade in superfluid turbulence},}\ }\href {\doibase
  10.1103/PhysRevLett.86.3080} {\bibfield  {journal} {\bibinfo  {journal}
  {Phys. Rev. Lett.}\ }\textbf {\bibinfo {volume} {86}},\ \bibinfo {pages}
  {3080--3083} (\bibinfo {year} {2001})}\BibitemShut {NoStop}%
\bibitem [{\citenamefont {Chung}\ and\ \citenamefont
  {Zhang}(2009)}]{chung_prl103}%
  \BibitemOpen
  \bibfield  {author} {\bibinfo {author} {\bibfnamefont {S.~B.}\ \bibnamefont
  {Chung}}\ and\ \bibinfo {author} {\bibfnamefont {S.-C.}\ \bibnamefont
  {Zhang}},\ }\bibfield  {title} {\enquote {\bibinfo {title} {Detecting the
  majorana fermion surface state of $^3${He-B} through spin relaxation},}\
  }\href@noop {} {\bibfield  {journal} {\bibinfo  {journal} {Phys. Rev. Lett.}\
  }\textbf {\bibinfo {volume} {103}},\ \bibinfo {pages} {235301} (\bibinfo
  {year} {2009})}\BibitemShut {NoStop}%
\bibitem [{\citenamefont {Kreil}\ \emph
  {et~al.}(2019{\natexlab{b}})\citenamefont {Kreil}, \citenamefont {Pomyalov},
  \citenamefont {L'vov}, \citenamefont {Musiienko-Shmarova}, \citenamefont
  {Melkov}, \citenamefont {Serga},\ and\ \citenamefont
  {Hillebrands}}]{alex2019josephson}%
  \BibitemOpen
  \bibfield  {author} {\bibinfo {author} {\bibfnamefont {Alexander J.~E.}\
  \bibnamefont {Kreil}}, \bibinfo {author} {\bibfnamefont {Anna}\ \bibnamefont
  {Pomyalov}}, \bibinfo {author} {\bibfnamefont {Victor~S.}\ \bibnamefont
  {L'vov}}, \bibinfo {author} {\bibfnamefont {Halyna~Yu.}\ \bibnamefont
  {Musiienko-Shmarova}}, \bibinfo {author} {\bibfnamefont {Gennadii~A.}\
  \bibnamefont {Melkov}}, \bibinfo {author} {\bibfnamefont {Alexander~A.}\
  \bibnamefont {Serga}}, \ and\ \bibinfo {author} {\bibfnamefont {Burkard}\
  \bibnamefont {Hillebrands}},\ }\bibfield  {title} {\enquote {\bibinfo {title}
  {Josephson oscillations in a room-temperature {Bose-Einstein} magnon
  condensate},}\ }\href@noop {} {\bibfield  {journal} {\bibinfo  {journal}
  {arXiv:1911.07802}\ } (\bibinfo {year} {2019}{\natexlab{b}})}\BibitemShut
  {NoStop}%
\bibitem [{\citenamefont {Bozhko}\ \emph {et~al.}(2019)\citenamefont {Bozhko},
  \citenamefont {Kreil}, \citenamefont {Musiienko-Shmarova}, \citenamefont
  {Serga}, \citenamefont {Pomyalov}, \citenamefont {L’vov},\ and\
  \citenamefont {Hillebrands}}]{Bozhko2019}%
  \BibitemOpen
  \bibfield  {author} {\bibinfo {author} {\bibfnamefont {D.A.}\ \bibnamefont
  {Bozhko}}, \bibinfo {author} {\bibfnamefont {A.J.E.}\ \bibnamefont {Kreil}},
  \bibinfo {author} {\bibfnamefont {H.Y.}\ \bibnamefont {Musiienko-Shmarova}},
  \bibinfo {author} {\bibfnamefont {A.A.}\ \bibnamefont {Serga}}, \bibinfo
  {author} {\bibfnamefont {A.}~\bibnamefont {Pomyalov}}, \bibinfo {author}
  {\bibfnamefont {V.S.}\ \bibnamefont {L’vov}}, \ and\ \bibinfo {author}
  {\bibfnamefont {B.}~\bibnamefont {Hillebrands}},\ }\bibfield  {title}
  {\enquote {\bibinfo {title} {Bogoliubov waves and distant transport of magnon
  condensate at room temperature},}\ }\href {\doibase
  10.1038/s41467-019-10118-y} {\bibfield  {journal} {\bibinfo  {journal}
  {Nature Communications}\ }\textbf {\bibinfo {volume} {10}} (\bibinfo {year}
  {2019}),\ 10.1038/s41467-019-10118-y}\BibitemShut {NoStop}%
\bibitem [{\citenamefont {Kreil}\ \emph {et~al.}(2018)\citenamefont {Kreil},
  \citenamefont {Bozhko}, \citenamefont {Musiienko-Shmarova}, \citenamefont
  {Vasyuchka}, \citenamefont {L'vov}, \citenamefont {Pomyalov}, \citenamefont
  {Hillebrands},\ and\ \citenamefont {Serga}}]{PhysRevLett.121.077203}%
  \BibitemOpen
  \bibfield  {author} {\bibinfo {author} {\bibfnamefont {Alexander J.~E.}\
  \bibnamefont {Kreil}}, \bibinfo {author} {\bibfnamefont {Dmytro~A.}\
  \bibnamefont {Bozhko}}, \bibinfo {author} {\bibfnamefont {Halyna~Yu.}\
  \bibnamefont {Musiienko-Shmarova}}, \bibinfo {author} {\bibfnamefont
  {Vitaliy~I.}\ \bibnamefont {Vasyuchka}}, \bibinfo {author} {\bibfnamefont
  {Victor~S.}\ \bibnamefont {L'vov}}, \bibinfo {author} {\bibfnamefont {Anna}\
  \bibnamefont {Pomyalov}}, \bibinfo {author} {\bibfnamefont {Burkard}\
  \bibnamefont {Hillebrands}}, \ and\ \bibinfo {author} {\bibfnamefont
  {Alexander~A.}\ \bibnamefont {Serga}},\ }\bibfield  {title} {\enquote
  {\bibinfo {title} {From kinetic instability to {Bose-Einstein} condensation
  and magnon supercurrents},}\ }\href {\doibase 10.1103/PhysRevLett.121.077203}
  {\bibfield  {journal} {\bibinfo  {journal} {Phys. Rev. Lett.}\ }\textbf
  {\bibinfo {volume} {121}},\ \bibinfo {pages} {077203} (\bibinfo {year}
  {2018})}\BibitemShut {NoStop}%
\bibitem [{\citenamefont {Bozhko}\ \emph {et~al.}(2016)\citenamefont {Bozhko},
  \citenamefont {Serga}, \citenamefont {Clausen}, \citenamefont {Vasyuchka},
  \citenamefont {Heussner}, \citenamefont {Melkov}, \citenamefont {Pomyalov},
  \citenamefont {L'Vov},\ and\ \citenamefont {Hillebrands}}]{Bozhko20161057}%
  \BibitemOpen
  \bibfield  {author} {\bibinfo {author} {\bibfnamefont {D.A.}\ \bibnamefont
  {Bozhko}}, \bibinfo {author} {\bibfnamefont {A.A.}\ \bibnamefont {Serga}},
  \bibinfo {author} {\bibfnamefont {P.}~\bibnamefont {Clausen}}, \bibinfo
  {author} {\bibfnamefont {V.I.}\ \bibnamefont {Vasyuchka}}, \bibinfo {author}
  {\bibfnamefont {F.}~\bibnamefont {Heussner}}, \bibinfo {author}
  {\bibfnamefont {G.A.}\ \bibnamefont {Melkov}}, \bibinfo {author}
  {\bibfnamefont {A.}~\bibnamefont {Pomyalov}}, \bibinfo {author}
  {\bibfnamefont {V.S.}\ \bibnamefont {L'Vov}}, \ and\ \bibinfo {author}
  {\bibfnamefont {B.}~\bibnamefont {Hillebrands}},\ }\bibfield  {title}
  {\enquote {\bibinfo {title} {Supercurrent in a room-temperature
  {Bose-Einstein} magnon condensate},}\ }\href {\doibase 10.1038/nphys3838}
  {\bibfield  {journal} {\bibinfo  {journal} {Nature Physics}\ }\textbf
  {\bibinfo {volume} {12}},\ \bibinfo {pages} {1057--1062} (\bibinfo {year}
  {2016})}\BibitemShut {NoStop}%
\end{thebibliography}
\end{document}